\documentclass[fleqn,usenatbib]{mnras}

\usepackage{newtxtext,newtxmath}
\usepackage[T1]{fontenc}

\DeclareRobustCommand{\VAN}[3]{#2}
\let\VANthebibliography\thebibliography
\def\thebibliography{\DeclareRobustCommand{\VAN}[3]{##3}\VANthebibliography}

\usepackage{graphicx}	% Including figure files
\usepackage{amsmath}	% Advanced maths commands
\newcommand{\TableNewLine}[2][c]{\begin{tabular}[#1]{@{}c@{}}#2\end{tabular}}
\usepackage{orcidlink}

\title[Sr Excitation Data for KNe \& WD Diagnostics]{Strontium {\sc i}, {\sc iii}, {\sc iv} and {\sc v}: Electron Impact Excitation Data for Kilonovae and White Dwarf Diagnostic Applications}

\author[D. J. Dougan et al.]{
David J. Dougan,$^{1}$\thanks{E-mail: ddougan04@qub.ac.uk}\orcidlink{0009-0008-3312-7426}
Niall E. McElroy,$^{1}$\orcidlink{0009-0002-7315-5444}
Connor P. Ballance$^{1}$\orcidlink{0000-0003-1693-1793}
and Catherine A. Ramsbottom$^{1}$\orcidlink{0000-0003-1579-8556}
\\
$^{1}$Astrophysics Research Centre, Queen’s University Belfast, Belfast, BT7 1NN, Northern Ireland, United Kingdom\\
}

\date{Accepted XXX. Received YYY; in original form ZZZ}

\pubyear{\the\year{}}

\begin{document}
\label{firstpage}
\pagerange{\pageref{firstpage}--\pageref{lastpage}}
\maketitle

\begin{abstract}
Strontium (Sr) emissions have been observed across a wide range of astrophysical phenomena, from kilonovae (KNe) events to white dwarf (WD) stars.  Precise and extensive atomic data for low ionisation stages of Sr is required for accurate theoretical modelling and to improve our understanding of evolutionary pathways.  We calculated energy levels, Einstein A coefficients and electron-impact excitation collision strengths for Sr {\sc i}, Sr {\sc iii}, Sr {\sc iv} and Sr {\sc v} at the temperature and density ranges of interest in KNe and WD research.  We developed new target structures using the {\sc grasp$^0$} and {\sc autostructure} packages.  The energies and A-values arising from the new structures were found to be in good agreement with experimental and theoretical equivalents reported in the literature.  Maxwellian averaged electron impact collision strengths were calculated using the $R$-matrix approach, as applied through the {\sc darc} and {\sc rmbp} coding packages.  These are presented in adf04 file format.  The new data sets allowed us to construct synthetic spectra for the first five ionisation stages of Sr and probe possible density and temperature diagnostic lines.  The synthetic spectra within the KNe regime revealed possible Sr {\sc iv} and Sr {\sc v} candidate lines at 1027.69nm and 1203.35nm respectively.  These may provide useful benchmarks for determining the extent of Sr ionisation that can be reached in an evolving KNe event.  Additional diagnostic lines were found to be poor across the Sr ion stages for both KNe and WD regimes due to most levels being in either coronal or Local Thermodynamic Equilibrium (LTE) conditions.
\end{abstract}


%%%%%%%%%%%%%%%%%%%%%%%%%%%%%%%%%%%%%%%%%%%%%%%%%%

%%%%%%%%%%%%%%%%% BODY OF PAPER %%%%%%%%%%%%%%%%%%

\section{Introduction}

%Sr ($Z = 38$) is an alkaline-earth element discovered in 1790 (\cite{Crawford_1790}) and named after the village in the Scottish Highlands where it was first mined (\cite{Partington_1981}).} 
Emission lines from low ionisation stages of Sr have been observed, identified and studied across a wide range of different laboratory and astrophysical related disciplines.  Neutral Sr, with its completely closed electron shell ground state, and its small collisional cross section (\cite{Ferrari_2006}), is a promising candidate in the development of atomic clocks.  A Sr lattice clock utilising the 5s5p ($^3$P$_0$) $\rightarrow$ 5s$^2$  ($^1$S$_0$) magnetic quadrupole transition as the 'pendulum', results in timekeeping to an extremely low degree of uncertainty (< 1x10$^{-16}$s) (\cite{Falke_2014}, \cite{Aeppli_2024}).  Other transitions, such as the 5s5p ($^3$P$_2$) $\rightarrow$ 5s$^2$ ($^1$S$_0$) are also being explored (\cite{Trautmann_2023}, \cite{Klusener2024}). Much work has been undertaken in cooling these Sr atoms to < 1$\mu$K temperatures required to minimise perturbation effects during the operation of these atomic clocks (\cite{Sorrentino_2006}, \cite{Qiao_2019}).

Within an astrophysical context, Sr lines from both the neutral and subsequent ion stages have been observed in many distant astronomical phenomena. The majority of these spectral lines have been for Sr {\sc ii}, but weak Sr {\sc i} lines have also been detected in the solar spectrum (\cite{Sullivan_1938}), as well as proposed Sr {\sc v} lines.  \cite{Barklem_2000} demonstrated that there may be broadening effects present in these Sr lines.  A broad P Cygni spectral line of Sr {\sc ii} was detected in the KNe event AT2017gfo from the merger of two neutron stars (\cite{watson_2019}).  Sr {\sc ii} lines are typically shown to be in very high abundance in R Coronae Borealis (RCB) and dustless Hydrogen-deficient Carbon Stars (dLHdC).  Such stars are found to have very high abundances of Sr {\sc ii}, especially at 4077\AA \space and 4215\AA ~(\cite{Vanture_1999}, \cite{Goswami_2013}, \cite{Crawford_2022}).  These strong Sr {\sc ii} lines can be used to measure the relative stellar abundances of the elements within stars (\cite{bergemann_2012}, \cite{Aoki_2022}). The 10327\AA \space and 10915\AA \space lines may also be used to determine stellar abundances, as was shown in \cite{Andrievsky_2011}.  The temperatures and densities of these distant bodies mean that there are low abundances of Sr {\sc i} present at the sources of these distant objects, and this results in very weak Sr {\sc i} lines being observed.  However, these lines, particularly the 4607.34\AA \space line, may still be used to measure stellar abundances, though most comparisons will employ both Sr {\sc i} and Sr {\sc ii} lines (for example see \cite{Gratton_1994}, \cite{Cowan_2002}, \cite{bergemann_2012}).  \cite{Rauch_2017} proposed the existence of numerous Sr {\sc v} lines in the spectra of the hot WD star RE 0503-289.  Sr rich sources have also been observed in distant dwarf galaxies like Draco (\cite{Cohen_2009}) and Canes Venatici {\sc i} \& {\sc ii} (\cite{Francois_2016}), as well as early-type galaxies (\cite{Conroy_2013}).

These Sr-rich sources are thought to originate through the r- and s- processes during the formation of elements heavier than iron.  The successive neutron capture and $\beta^{-}$ decay are differentiated based on the timescale of the decay to the capture, and are believed to determine the origin and relative abundances of these heavier elements (\cite{Kasen_2017}), where Sr forms the first peak in this formation chain along with yttrium (Y) and zirconium (Zr) (\cite{Sneppen_2023}, \cite{Vieira_2023}, \cite{Hotokezaka_2023}).  These processes can only occur where there is a sufficiently high density of free neutrons. The violent interactions between compact binary systems (\cite{Lattimer_1977}) and the evolution of stars into the Asymptotic Giant Branch (AGB) (\cite{Smith_1990}) are believed to provide the necessary conditions for the r- and s-processes to occur and form Sr.  However, there may be an additional reaction pathway, as particular strong features cannot be adequately explained by a pure r-/s-process alone (\cite{Hirai_2019}).       

Computational modelling of these astrophysical phenomena yields a greater understanding of the evolution pathway during their life cycle.  With sophisticated radiative transfer packages such as {\sc tardis} (\cite{Kerzendorf_2014}, \cite{Vogl_2019}), {\sc cloudy} (\cite{Ferland_2017}) and {\sc colradpy} (\cite{Johnson_2019}), synthetic spectra and the diagnostic lines that can arise from these astrophysical events in both LTE and Non-LTE (NLTE) conditions may be generated. These programs require input regarding the various species which form the basis of the synthetic spectra. Sr can play a notable role in these models, from KNe events (\cite{Gillanders_2022}, (\cite{Vieira_2024})), to galaxy evolution (\cite{Hirai_2019}). There is discussion about whether the P Cygni line detected in the KNe AT2017gfo event is Sr or if it is the result of a lighter element such as helium (\cite{Perego_2022}). Computational modelling can help us to discern which is the more probable origin (\cite{Tarumi_2023}).    

How the production of Sr affects the evolution pathway of these astrophysical phenomena requires a comprehensive understanding of radiative data, such as energy levels, both in the ground state and subsequent ionised stages, and the resulting transitions arising from each structure. This may include the collisional processes of electron-impact excitation, ionisation, photoionisation and recombination. The recent technical advancements and wider availability of computational resources has allowed collisional radiative solver packages to accept and process more extensive and comprehensive atomic data sets for use in modelling.  However, the atomic data generation itself has not kept pace, and as a result, many gaps remain in the datasets. The energy levels for each Sr species are well established, and are compiled and presented in the National Institute of Standards and Technology (NIST, \cite{NIST_2023}).  This has allowed structure models for Sr {\sc i} (\cite{Hassouneh_2022}), Sr {\sc ii} (\cite{mulholland_2024}), Sr {\sc iv} (\cite{Aggarwal_2015}, \cite{Rauch_2017}), Sr {\sc v} (\cite{Rauch_2017}, \cite{Aloui_2022}) and Sr {\sc vi} (\cite{Aloui_2022}) to be constructed and the Einstein A coefficients for spontaneous emission to be determined.  In contrast, much work remains in determining the other rates.  The photoionisation rates of Sr {\sc i} were determined in \cite{bergemann_2012} and \cite{fernández-menchero_2020}.  Electron impact excitation rates for Sr {\sc ii} were determined in \cite{Bautista_2002} and more recently by \cite{mulholland_2024}. There is limited published data for the other Sr low ionised species of interest highlighted in this paper. The electron-impact excitation rates can be approximated using the \cite{vanRegemorter_1962} and \cite{Axelrod_1980} techniques for allowed and forbidden transitions respectively.  However, the work of \cite{mulholland_2024} shows that there can be discrepancies between these approximations when compared with other more complete methods, such as the $R$-matrix method. This is especially true for the forbidden transitions with the \cite{Axelrod_1980} approach.  

%We aim in this paper to supplement the available Sr data available for astrophysical modelling.  The initial aim is to focus on the first five ionized stages of strontium, from Sr I to Sr V, starting with the electron-impact excitation rates.  Our main motivation for preparing these data sets is the modelling of kilonovae events, but the data is intended to be multi-purpose and may be used for both astrophysical and laboratory based applications.  It is hoped that the calculation and preparation of these strontium data sets at the same time will ensure a consistency across the data sets.  The subsequent data sets will be available on the \cite{OPEN_ADAS} website.  

The subsequent sections of this paper are structured as follows.  \text{Section \ref{sec:Atomic-Structure}} is an overview of the atomic structure theories used to generate the atomic models for Sr {\sc i}, {\sc iii}, {\sc iv} and {\sc v}. A comprehensive comparison of the energy levels and spontaneous emission rates arising from each model is made with all available data in the literature.  In \text{Section \ref{sec:Electron-Impact-Excitation}} the collisional theory and codes used to generate both the collision strengths and the Maxwellian averaged effective collision strengths are discussed. A selection of results for some representative transitions in each species are presented. In \text{Section \ref{sec:Colisional Radiative Modelling}}, the atomic data are incorporated into a full NLTE collisional radiative solver. Possible temperature and density diagnostics are identified through investigations of the level populations of each species. This will culminate in a complete synthetic spectra consisting of the first five ion stages of strontium at temperatures and densities suitable for KNe and WD research.  \text{Section \ref{sec:Conclusions}} summarises and concludes our findings.

\section{Atomic Structure - S\MakeLowercase{r} Models}
\label{sec:Atomic-Structure}

\subsection{Methodology}

 Electron-impact excitation rates are underpinned by the accuracy of the atomic structure. To generate accurate atomic structure models, we employ the use of two different, but well-established  atomic structure codes, the semi-relativistic {\sc autostructure} package and the fully relativistic General Relativistic Atomic Structure Package {\sc grasp$^0$}. A brief summary of the methodology behind these packages is provided below.

The {\sc autostructure} (AS) package is a successor to the {\sc superstructure} package developed by \cite{eissner_1991} and has been heavily revised by \cite{badnell_1986} and \cite{badnell_1997}.  The atomic structure of the target is constructed through solving the time independent Schr\"{o}dinger equation by considering a semi-relativistic N-electron Breit-Pauli Hamiltonian ($H_{BP}$) given by

\begin{equation} \label{Eq_Autostructure_1}
    H_{BP}\phi = E\phi
\end{equation}

where $\phi$ are the target wavefunctions, and $E$ the energy eigenvalues. $H_{BP}$ may be broken down into the non-relativistic and relativistic constituent operators (in atomic units) as follows
\begin{equation} \label{Eq_Autostructure_2}
    H_{BP} = H_{NR} + H_{RC}
\end{equation}

with $H_{NR}$ including the non-relativistic operators so that
\begin{equation} \label{Eq_Autostructure_3}
    H_{NR} = \sum_{i=1}^N{\left(-\frac{1}{2}\bigtriangledown^2_i - \frac{Z}{r_i}\right)} + \sum_{i>j=1}^N{\frac{1}{r_{ij}}}.
\end{equation}

and $H_{RC}$ including the first order mass, Darwin and spin-orbit relativistic correction operators.

\begin{equation} 
\label{Eq_Autostructure_4}
    H_{RC} = \frac{\alpha^2 Z}{2} \sum_{i=1}^N{\frac{\mathbf{l_i . s_i}}{r^3_i}} - \frac{\alpha^2}{8} \sum_{i=1}^N{\bigtriangledown^4_i} - \frac{\alpha^2Z}{8} \sum^N_{i=1}{\bigtriangledown^2_i}\left(\frac{1}{r_i}\right)
\end{equation}

In Eqn. \ref{Eq_Autostructure_4}, $\mathbf{l_i}$ is the single electron orbital operator, $\mathbf{s_i}$ is the single electron spin angular momentum operator, $Z$ is the atomic number of the target, $r_{i}$ is the position of electron $i$ from the target centre, and $r_{ij}$ is the interelectronic distance ($| r_i - r_j|$).
The parameters for each orbital ($nl$) are generated through consideration of a Thomas–Fermi–Dirac–Amaldi potential model.  These can be further modified though the use of scaling factors ($\lambda_{nl}$) on each orbital.  Each $\lambda_{nl}$ is varied such that the orbitals generated minimise the energy of the Hamiltonian and these target energies mimimise the absolute 
difference with NIST values (\cite{NIST_2023}).  

{\sc grasp}$^0$ is based on the Multi-Configuration Dirac–Fock (MCDF) codes developed by \cite{grant_1984} and \cite{mckenzie_1984} and later published in \cite{dyall_1989}.  We chose a Dirac-Columb Hamiltonian ($H_D$) defined as     

\begin{equation} \label{Eq_GRASP_2}
    H_D = \sum_{i=1}^N\left({c\mathbf{\alpha . p_i}} + \left(\beta - I_4\right)^2 - \frac{Z}{r_i}\right) + \sum^N_{i>j=1}{\frac{1}{r_{ij}}}
\end{equation}

where $c$ is the speed of light in a vacuum, $\alpha$ and $\beta$ are the Pauli spin matrices, $I_4$ is the (4 x 4) identity matrix, and $\mathbf{p_i}$ is the momentum operator where $\mathbf{p_i} = -i\bigtriangledown$. {\sc grasp}$^0$ determines and optimises the orbitals of the target species by minimising the energy associated with $H_D$.

In this work we have created four new fine-structure resolved models for Sr {\sc i}, Sr {\sc iii}, Sr {\sc iv} and Sr {\sc v}.  The singly ionised Sr {\sc ii} ion is not included in this list as it has already been considered in detail by \cite{mulholland_2024}. A summary of the four models is provided in \text{Table \ref{tab:Structure_Summary}} where it is evident that for some charge states the optimum model was obtained using the {\sc grasp}$^{0}$ package and for the remainder {\sc autostructure} provided the best model. A more detailed description of each target structure is given below. 

\begin{table}
\centering
\caption[]{Summary of Sr Atomic Structures.}
\begin{tabular}{ccccc}\hline
\textbf{Species} & \TableNewLine{\textbf{Structure} \\ \textbf{Package}} & \TableNewLine{\textbf{Num.} \\ \textbf{CSFs}} & \TableNewLine{\textbf{Num.} \\ \textbf{Orbitals}} & \TableNewLine{\textbf{Num.} \\ \textbf{Levels}} \\
\hline
\noalign{\smallskip}
Sr {\sc i} & AS & 29 & 19 & 202 \\ \noalign{\smallskip}
Sr {\sc iii} & {\sc grasp}$^0$ & 24 & 20 & 853 \\ \noalign{\smallskip}
Sr {\sc iv} & {\sc grasp}$^0$ & 21 & 19 & 504 \\ \noalign{\smallskip}
Sr {\sc v} & AS & 23 & 20 & 1,817 \\ \noalign{\smallskip}
\hline
\noalign{\smallskip}
\end{tabular}
\label{tab:Structure_Summary}
%}
\end{table}

\subsection{Sr {\sc i} Model} \label{sec:SrI_Structure}

Neutral strontium consists of 38 electrons, with a ground state of [Kr] 5s$^2$ ($^1$S$_0$), making it analogous to a helium atom.  With a relatively low ionisation potential of 0.55~Ryd (7.50~eV), there will typically be only trace amounts of Sr {\sc i} present in most observed astrophysical plasmas, and as such observed Sr {\sc i} lines tend to be very weak.  The AS model of Sr {\sc i} was constructed using 19 orbitals extending to $n$=8 and $l$=2.  They were 1s, 2s, 3s, 3p, 3d, 4s, 4p, 4d, 5s, 5p, 5d, 6s, 6p, 6d, 7s, 7p and 8s.  It comprised 29 configurations, listed in \text{Table \ref{tab:Sr_I_Structure}}, the principal ones including a single electron promotion out of the 5s$^2$ orbital in the ground configuration into the higher energy orbitals.  The resulting target consists of 202 fine-structure levels in total. 
The finalised $\lambda_{nl}$ parameters adopted in the computations are displayed in \text{Table \ref{tab:Sr_I_lambda}}. 

   \begin{table}
   
     \centering
     \caption[]{The configurations included in the wavefunction expansion for the AS model of Sr {\sc i}.}
         \begin{tabular}{p{2cm} p{2cm } p{2cm}}
            \hline
            \noalign{\smallskip}
            \multicolumn{3}{l}{\textbf{Sr {\sc i} - 29 Configurations}} \\
            \noalign{\smallskip}
            \hline
            \noalign{\smallskip}
            4p$^{6}$4d5s   & 4p$^{6}$4d5p  & 
            4p$^{6}$4d5d  \\ \noalign{\smallskip}
            4p$^{6}$4d6s   & 4p$^{6}$4d6p    & 4p$^{6}$4d6d  \\ \noalign{\smallskip}
            4p$^{6}$4d7s   & 4p$^{6}$4d7p    & 4p$^{6}$4d8s \\ \noalign{\smallskip}
            4p$^{6}$4f5s   & 4p$^{6}$5s5p    & 4p$^{6}$5s5d  \\ \noalign{\smallskip}
            4p$^{6}$5s6s   & 4p$^{6}$5s6p    & 4p$^{6}$5s6d  \\ \noalign{\smallskip}
            4p$^{6}$5s7s   & 4p$^{6}$5s7p    & 4p$^{6}$5s8s \\ \noalign{\smallskip}
            4p$^{6}$4d$^{2}$   & 4p$^{6}$4f$^{2}$    & 4p$^{6}$5s$^{2}$  \\ \noalign{\smallskip}
            4p$^{6}$5p$^{2}$   & 4p$^{6}$5p5d    & 4p$^{6}$5p6s \\ \noalign{\smallskip}
            4p$^{6}$5p6p   & 4p$^{6}$5p6d    & 4p$^{6}$5p7s \\ \noalign{\smallskip}
            4p$^{6}$5p7p   & 4p$^{6}$5p8s    & {} \\ \noalign{\smallskip}
            \hline
            \noalign{\smallskip}
         \end{tabular}

        \label{tab:Sr_I_Structure}
   \end{table}

   \begin{table}
     \centering
        \caption[]{The $\lambda_{nl}$ scaling parameters used in the Sr {\sc i} model for each orbital.}
         \begin{tabular}{p{2.0cm}p{2.0cm}p{2.0cm}}
            \hline
            \noalign{\smallskip}
            \multicolumn{3}{l}{\textbf{Sr {\sc i} - $\lambda_{nl}$ Values}} \\
            \noalign{\smallskip}
            \hline
            \noalign{\smallskip}
             1s - 0.9171 & 2s - 0.9353 & 2p - 1.0584  \\ \noalign{\smallskip}
             3s - 0.9494 & 3p - 0.9969 & 3d - 1.2457  \\ \noalign{\smallskip}
             4s - 1.1023 & 4p - 1.0005 & 4d - 1.0343  \\ \noalign{\smallskip}
             4f - 0.8322 & 5s - 1.0562 & 5p - 1.0210  \\ \noalign{\smallskip}
             5d - 1.1818 & 6s - 0.8707 & 6p - 1.1216  \\ \noalign{\smallskip}
             6d - 1.2266 & 7s - 1.2289 & 7p - 1.2490  \\ \noalign{\smallskip}
             8s - 0.8397 & {} & {}  \\ \noalign{\smallskip}
            \hline
            \noalign{\smallskip}
         \end{tabular}
        
        \label{tab:Sr_I_lambda}
   \end{table}

The NIST database consists of 380 experimental energy levels for Sr {\sc i}, obtained from the works of \cite{Garton_1968}, \cite{Hudson_1969}, \cite{Newsom_1973}, \cite{Rubbmark_1978}, \cite{Armstrong_1979}, \cite{Beigang_1982_b}, \cite{Beigang_1982_a}, \cite{Kompitsas_1990}, \cite{Kompitsas_1991}, \cite{Goutis_1992}, \cite{Jimoyiannis_1993}, \cite{Kunze_1993}, \cite{Dai_1995}, \cite{Dai_1996} and \cite{sansonetti_2010}.  
In total, we were able to identify the lowest 57 levels between the AS target and NIST, covering the energy range from 0.00 -$~$0.45 Ryd. In \text{Table \ref{tab:Sr_I_Energy_Levels}} we present a comparison of the lowest lying 30 of these energy levels and  very good agreement is found with those reported in NIST. The levels deviate by $<$0.01 Ryd and the average relative percentage difference was found to be -2.10\%.

\begin{table}
\centering
\caption[]{The first 30 energy levels of the Sr {\sc i} model. The NIST values for Levels 1 - 29 were obtained from \cite{sansonetti_2010}, with level 30 from \cite{Dai_1996} and \cite{sansonetti_2010}.}
%\resizebox{2\columnwidth}{!}{%
\begin{tabular}{ccccc}\hline
\textbf{Level} & \textbf{Config.}      & \TableNewLine{\textbf{AS} \\ \textbf{/ Ryd}} & \TableNewLine{\textbf{NIST} \\ \textbf{/ Ryd}} & \TableNewLine{\textbf{Percentage} \\ \textbf{Error / \%}} \\\hline
\noalign{\smallskip}
1     & 5s$^2$\,($^1$S$_0$)         & 0.00000             & 0.00000                 &                     \\  \noalign{\smallskip}
2     & 5s5p\,($^3$P$^{\circ}_0$)   & 0.12958             & 0.13047               & -0.681             \\ \noalign{\smallskip}
3     & 5s5p $(^3$P$^{\circ}_1)$      & 0.13126             & 0.13217               & -0.692             \\ \noalign{\smallskip}
4     & 5s5p $(^3$P$^{\circ}_2)$      & 0.13475             & 0.13577               & -0.749             \\ \noalign{\smallskip}
5     & 4d5s $(^3$D$_1)        $      & 0.16549             & 0.16548               & \phantom{-}0.005              \\ \noalign{\smallskip}
6     & 4d5s $(^3$D$_2)        $      & 0.16595             & 0.16602               & -0.043             \\ \noalign{\smallskip}
7     & 4d5s $(^3$D$_3)        $      & 0.16666             & 0.16694               & -0.165             \\ \noalign{\smallskip}
8     & 4d5s $(^1$D$_2)        $      & 0.19017             & 0.18362               & \phantom{-}3.570              \\ \noalign{\smallskip}
9     & 5s5p $(^1$P$^{\circ}_1)$      & 0.19856             & 0.19773               & \phantom{-}0.417              \\ \noalign{\smallskip}
10    & 5s6s $(^3$S$_1)        $      & 0.26458             & 0.26462               & -0.016             \\ \noalign{\smallskip}
11    & 5s6s $(^1$S$_0)        $      & 0.28700             & 0.27877               & \phantom{-}2.952              \\ \noalign{\smallskip}
12    & 4d5p $(^3$F$^{\circ}_2)$      & 0.30315             & 0.30315               & \phantom{-}0.000              \\ \noalign{\smallskip}
13    & 4d5p $(^3$F$^{\circ}_3)$      & 0.30582             & 0.30609               & -0.088             \\ \noalign{\smallskip}
14    & 4d5p $(^1$D$^{\circ}_2)$      & 0.30726             & 0.30825               & -0.322             \\ \noalign{\smallskip}
15    & 5s6p $(^3$P$^{\circ}_0)$      & 0.30830             & 0.30850               & -0.063             \\ \noalign{\smallskip}
16    & 5s6p $(^3$P$^{\circ}_1)$      & 0.30840             & 0.30863               & -0.075             \\ \noalign{\smallskip}
17    & 4d5p $(^3$F$^{\circ}_4)$      & 0.30844             & 0.30910               & -0.213             \\ \noalign{\smallskip}
18    & 5s6p $(^3$P$^{\circ}_2)$      & 0.30963             & 0.30959               & \phantom{-}0.015              \\ \noalign{\smallskip}
19    & 5s6p $(^1$P$^{\circ}_1)$      & 0.31042             & 0.31073               & -0.098             \\ \noalign{\smallskip}
20    & 5s5d $(^1$D$_2)        $      & 0.31689             & 0.31646               & \phantom{-}0.137              \\ \noalign{\smallskip}
21    & 5s5d $(^3$D$_1)        $      & 0.31881             & 0.31901               & -0.063             \\ \noalign{\smallskip}
22    & 5s5d $(^3$D$_2)        $      & 0.31895             & 0.31914               & -0.060             \\ \noalign{\smallskip}
23    & 5s5d $(^3$D$_3)        $      & 0.31916             & 0.31935               & -0.060             \\ \noalign{\smallskip}
24    & 5p$^2$ $(^3$P$_0)        $    & 0.32690             & 0.32071               & \phantom{-}1.931              \\ \noalign{\smallskip}
25    & 5$p^2$ $(^3$P$_1)        $    & 0.32842             & 0.32259               & \phantom{-}1.808              \\ \noalign{\smallskip}
26    & 5$p^2$ $(^3$P$_2)        $    & 0.33046             & 0.32509               & \phantom{-}1.652              \\ \noalign{\smallskip}
27    & 4d5p $(^3$D$^{\circ}_1)$      & 0.33372             & 0.33046               & \phantom{-}0.984              \\ \noalign{\smallskip}
28    & 4d5p $(^3$D$^{\circ}_2)$      & 0.33462             & 0.33153               & \phantom{-}0.930              \\ \noalign{\smallskip}
29    & 4d5p $(^3$D$^{\circ}_3)$      & 0.33604             & 0.33315               & \phantom{-}0.866              \\ \noalign{\smallskip}
30    & 4d$^2$ $(^1$D$_2)        $    & 0.33849             & 0.33681               & \phantom{-}0.498             \\ \noalign{\smallskip}
\hline
\noalign{\smallskip}
\end{tabular}%

\label{tab:Sr_I_Energy_Levels}
%}
\end{table}

\begin{figure}
\centering

\includegraphics[width=1.0\columnwidth]{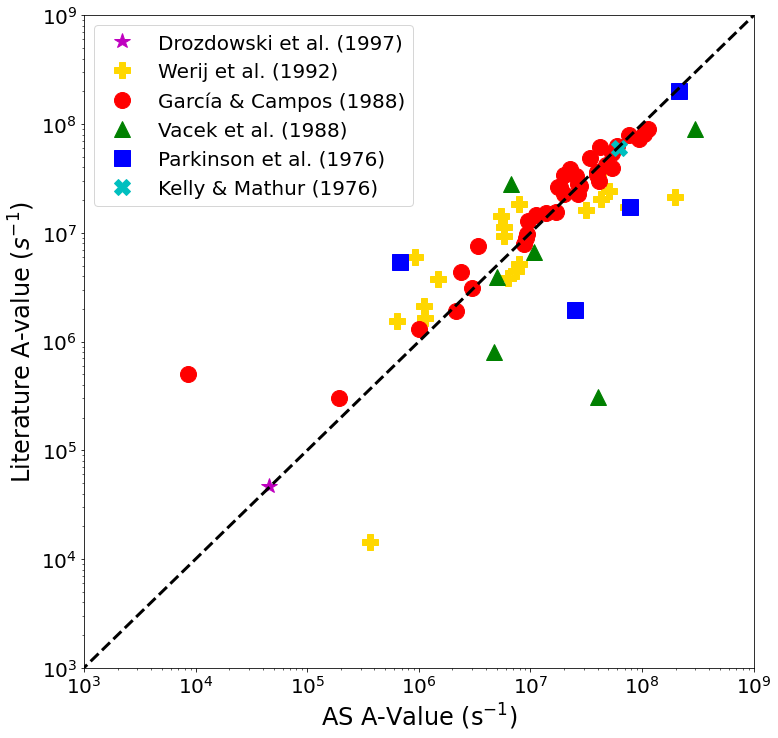}
\caption{A comparison of the Einstein A-values for dipole transitions from the AS Sr {\sc i} model and those for which there is an equivalent literature value.  The literature A-values were obtained from \protect\cite{Kelly_1976}, \protect\cite{Parkinson_1976}, \protect\cite{Vacek_1988}, \protect\cite{Garcia_1988}, \protect\cite{Werij_1992} and \protect\cite{Drozdowski_1997}.}
\label{fig:SrI_A-values}.
\end{figure}

A second test of the accuracy of this Sr {\sc i} target can be achieved by comparing the computed Einstein A-coefficients with experimentally measured values. It is important to note that the transition rates are heavily dependent on the energy separation between the initial and final levels ($\Delta E$).  It scales by ($\Delta E$)$^3$ for electric dipole (E1) transitions, ($\Delta E$)$^5$ for electric quadrupole (E2) and magnetic dipole (M1) transitions, and ($\Delta E$)$^7$ for electric octupole (E3) and magnetic quadrupole (M2) transitions.  To accommodate for this, the A-values presented in this section were calculated using the spectroscopically accurate measured data reported in NIST.  The A-values were rescaled as follows

\begin{equation} \label{A_value_Shift}
    A_{j \rightarrow i}(\textrm{Shifted}) = \Big(\frac{\lambda_{\textrm{Calculated}}}{\lambda_{\textrm{NIST}}}\Big)^m A_{j \rightarrow i}(\textrm{Unshifted})
\end{equation}

where $m$=3 for E1 dipole transitions, 
$m$=5 for E2/M1 transitions, and $m$=7 for E3/M2 transitions. The A-values listed in NIST are from the compilation work of \cite{sansonetti_2010}, which presents transition rates from \cite{Kelly_1976}, \cite{Parkinson_1976}, \cite{Garton_1983}, \cite{Garcia_1988}, \cite{Vacek_1988}, \cite{Connerade_1992}, \cite{Werij_1992} and  \cite{Drozdowski_1997}.  All these transitions are classified as electric dipole. In \text{Table \ref{tab:SrI-A-values}} we present a comparison of a selection of the strongest lines. In particular, we highlight the agreement between the 460.773nm and 707.001nm Sr {\sc i} lines which, as discussed in \cite{bergemann_2012}, are useful in determining the stellar abundances of stars approaching the end of their lives. We also compare with the Fock-space relativistic coupled-cluster (FSRCC) calculations of \cite{Patel_2024} relevant in pumping applications, and the experimental and theoretical work of \cite{Klusener2024} for use in probing probable transitions for atomic clocks.  There is good agreement between the literature A-values and our calculated values, with the majority within the same order of magnitude.  There is a prominent exception with the (1-4) M2 transition where there is a two order of magnitude difference between the AS A-value and that determined in  \cite{Klusener2024}.  It is generally more challenging to achieve convergence for these weaker M2 quadrupole lines.
In \text{Figure \ref{fig:SrI_A-values}} we present a graphical summary of these comparisons for the strongest lines with A-values between 10$^3$ and 10$^9$~s$^{-1}$, the dashed line representing the line of equality between the current predictions and those from the literature. There is good clustering around this line especially for the stronger E1 transitions.  The weaker lines exhibit a greater degree of variation.

\begin{table}

  \centering
  \caption{Comparison between sample Einstein A-values obtained from the Sr {\sc i} AS target and those obtained from [1]-\protect\cite{Klusener2024}, [2]-\protect\cite{Parkinson_1976}, [3]-\protect\cite{Garcia_1988}, [4]-\protect\cite{Patel_2024} and [5]-\protect\cite{Werij_1992}.  The Index column refers to the energy levels displayed in \text{Table \ref{tab:Sr_I_Structure}}. The wavelengths corresponding to each transition were experimentally measured in \protect\cite{Meggers_1933} and \protect\cite{Sullivan_1938}.}
    \begin{tabular}{cccc} \hline
            \textbf{Index} & \TableNewLine{\textbf{Wavelength} \\ \textbf{ / nm}}      & \TableNewLine{\textbf{AS} \\ \textbf{A-value / s$^{-1}$} } & \TableNewLine{\textbf{Literature}\\ \textbf{A-value / s$^{-1}$} }  \\ \hline
 \noalign{\smallskip}
    1 - 4\phantom{-} & 672.9613 & 1.18E-04 & 1.12E-06$^{[1]}$ \\ \noalign{\smallskip}
    1 - 9\phantom{-} & 460.7730 & 2.10E+08 & 2.13E+08$^{[2]}$ \\ \noalign{\smallskip}
    2 - 21     & 483.2043 & 2.58E+07 & 3.32E+07$^{[3]}$ \\ \noalign{\smallskip}
     & & & 3.52E+07$^{[4]}$ \\ \noalign{\smallskip}
    3 - 10     & 687.8313 & 2.79E+07 & 2.72E+07$^{[3]}$ \\ \noalign{\smallskip}
    3 - 21     & 487.6075 & 1.78E+07 & 2.63E+07$^{[3]}$ \\ \noalign{\smallskip}
    & & & 3.29E+07$^{[4]}$ \\ \noalign{\smallskip}
    4 - 10     & 707.0010 & 4.63E+07 & 4.74E+07$^{[3]}$ \\ \noalign{\smallskip}
    4 - 21     & 497.1668 & 9.94E+05 & 1.31E+06$^{[3]}$ \\ \noalign{\smallskip}
    & & & 8.32E+05$^{[4]}$ \\ \noalign{\smallskip}
    5 - 12     & 661.7266 & 3.17E+07 & 1.63E+07$^{[5]}$ \\ \noalign{\smallskip}
    6 - 18     & 634.5726 & 1.11E+06 & 2.14E+06$^{[5]}$ \\ \noalign{\smallskip}
    6 - 28     & 550.4184 & 5.36E+07 & 5.44E+07$^{[3]}$ \\ \noalign{\smallskip}
    7 - 17     & 640.8463 & 5.07E+07 & 2.45E+07$^{[5]}$ \\ \noalign{\smallskip}
    9 - 21     & 753.3638 & 8.04E+03 & 5.20E+09$^{[4]}$  \\ \noalign{\smallskip} 
    \hline
    \noalign{\smallskip}
    \end{tabular}%
    \label{tab:SrI-A-values}%
\end{table}

\subsection{Sr {\sc iii} Model} \label{sec:SrIII_Model}

Sr {\sc iii} is part of the Krypton isoelectronic sequence, and as such has 36 electrons.  It has a ground state of [Ar] 3d$^{10}$4s$^2$4p$^6$ ($^1$S$_0$), making it analogous to a Noble gas in structure.  The present {\sc grasp$^0$} structure of Sr {\sc iii} consists of 20 orbitals up to $n$=7 and $l$=3 (1s, 2s, 2p, 3s, 3p, 3d, 4s, 4p, 4d, 4f, 5s, 5p, 5d, 5f, 6s, 6p, 6d, 6f, 7s and 7p). The 24 configurations included in the model are listed in \text{Table \ref{tab:Sr_III_Structure}} and a total of 853 even and odd levels were computed.

   \begin{table}
           \caption[]{The configurations included in the wavefunction expansion for the GRASP$^0$ model for Sr {\sc iii}.}
         \begin{tabular}{p{2.0cm}p{2.0cm}p{2.0cm}}
            \hline
            \noalign{\smallskip}
            \multicolumn{3}{l}{\textbf{Sr {\sc iii} - 24 Configurations}} \\
            \noalign{\smallskip}
            \hline
            \noalign{\smallskip}
            4p$^{6}$          &  4p$^{5}$4d      & 4p$^{5}$4f  \\ \noalign{\smallskip}
            4p$^{5}$5s      &  4p$^{5}$5p      & 4p$^{5}$5d  \\ \noalign{\smallskip}
            4p$^{5}$5f      &  4p$^{5}$6s      & 4p$^{5}$6p \\ \noalign{\smallskip}
            4p$^{5}$6d      &  4p$^{5}$6f      & 4p$^{5}$7s  \\ \noalign{\smallskip}
            4p$^{5}$7p      &  4p$^{4}$4d5s  & 4p$^{4}$4f5s  \\ \noalign{\smallskip}
            4p$^{4}$4f5p  &  4p$^{4}$5s5f  & 4p$^{4}$5s5d \\ \noalign{\smallskip}
            4p$^{4}$5p6s  &  4p$^{4}$5d6s  & 4p$^{4}$5s$^{2}$  \\ \noalign{\smallskip}
            4p$^{4}$5d$^{2}$  &  4p$^{4}$6s$^{2}$  & 4p$^{4}$6d$^{2}$ \\ \noalign{\smallskip}
            \hline
            \noalign{\smallskip}
         \end{tabular}
        \label{tab:Sr_III_Structure}
   \end{table}

The NIST database contains 150 energy levels for Sr {\sc iii}, obtained from \cite{Reader_1972}, \cite{Persson_1972}, \cite{Hansen_1973} and \cite{sansonetti_2012}.  It was possible to match the current 65 lowest lying energy levels included in the target up to an energy of 2.70 Ryd, which greatly exceeds the temperatures observed in KNe and WD applications.  A comparison between the first 30 energy levels is provided in \text{Table \ref{tab:SrIII-Energy-Levels}} where good agreement is evident, with differences of $<0.1$ Ryd for most levels.  The average relative percentage difference between the 65 shifted and unshifted energy levels is 2.618\%, with the difference not exceeding 4.05\% for any individual level.  It should be noted that due to the ground configuration of Sr {\sc iii} being a completely closed system, significant energy must be applied to bring the system to its lowest lying excited state ($\approx$1.60 Ryd).  This means that there are no Sr {\sc iii} levels lying within the observed KNe temperatures and hence it will be unlikely that Sr {\sc iii} emissions will be observed unless higher temperature sources are investigated.

\begin{table}
  \centering
    \caption{The first 30 energy levels in the Sr {\sc iii} model. The NIST energies for all of the first 30 energy levels were obtained from \protect\cite{Persson_1972} and \protect\cite{Hansen_1973}.}
    \begin{tabular}{ccccc} \hline
            \textbf{Level} & \textbf{Config.}      & \TableNewLine{\textbf{GRASP$^0$} \\ \textbf{/ Ryd}} & \TableNewLine{\textbf{NIST} \\ \textbf{/ Ryd}} & \TableNewLine{\textbf{Percentage} \\ \textbf{Error / \%}} \\ \hline
 \noalign{\smallskip}
    1     & 4p$^{6}$\, ($^{1}$S$_{0}$)          & 0.00000 & 0.00000 &            \\ \noalign{\smallskip}
    2     & 4p$^{5}$4d\, ($^{3}$P$_{0}^{\circ}$)   & 1.60015 & 1.60779 & \phantom{-}0.475 \\ \noalign{\smallskip}
    3     & 4p$^{5}$4d $(^{3}$P$_{1}^{\circ})$   & 1.61096 & 1.61931 & \phantom{-}0.515 \\ \noalign{\smallskip}
    4     & 4p$^{5}$4d $(^{3}$P$_{2}^{\circ})$   & 1.63145 & 1.64223 & \phantom{-}0.656 \\ \noalign{\smallskip}
    5     & 4p$^{5}$4d $(^{3}$F$_{4}^{\circ})$   & 1.66856 & 1.66874 & \phantom{-}0.011 \\ \noalign{\smallskip}
    6     & 4p$^{5}$4d $(^{3}$F$_{3}^{\circ})$   & 1.68930 & 1.68681 & -0.148 \\ \noalign{\smallskip}
    7     & 4p$^{5}$4d $(^{3}$F$_{2}^{\circ})$   & 1.71440 & 1.70922 & -0.303 \\ \noalign{\smallskip}
    8     & 4p$^{5}$5s $(^{3}$P$_{2}^{\circ})$   & 1.71825 & 1.75561 & \phantom{-}2.128 \\ \noalign{\smallskip}
    9     & 4p$^{5}$4d $(^{3}$D$_{3}^{\circ})$   & 1.77179 & 1.75971 & -0.686 \\ \noalign{\smallskip}
    10    & 4p$^{5}$5s $(^{1}$P$_{1}^{\circ})$   & 1.73094 & 1.77161 & \phantom{-}2.296 \\ \noalign{\smallskip}
    11    & 4p$^{5}$4d $(^{1}$D$_{2}^{\circ})$   & 1.80512 & 1.78873 & -0.916 \\ \noalign{\smallskip}
    12    & 4p$^{5}$4d $(^{3}$D$_{1}^{\circ})$   & 1.81462 & 1.79726 & -0.966 \\ \noalign{\smallskip}
    13    & 4p$^{5}$4d $(^{3}$D$_{2}^{\circ})$   & 1.82537 & 1.81104 & -0.792 \\ \noalign{\smallskip}
    14    & 4p$^{5}$4d $(^{1}$F$_{3}^{\circ})$   & 1.83213 & 1.82177 & -0.569 \\ \noalign{\smallskip}
    15    & 4p$^{5}$5s $(^{3}$P$_{0}^{\circ})$   & 1.80112 & 1.84150 & \phantom{-}2.193 \\ \noalign{\smallskip}
    16    & 4p$^{5}$5s $(^{3}$P$_{1}^{\circ})$   & 1.80874 & 1.85299 & \phantom{-}2.388 \\ \noalign{\smallskip}
    17    & 4p$^{5}$5p $(^{3}$S$_{1}) $          & 1.94569 & 2.02115 & \phantom{-}3.734 \\ \noalign{\smallskip}
    18    & 4p$^{5}$5p $(^{3}$D$_{2}) $          & 1.97474 & 2.05318 & \phantom{-}3.820 \\ \noalign{\smallskip}
    19    & 4p$^{5}$5p $(^{3}$D$_{3}) $          & 1.97795 & 2.05803 & \phantom{-}3.891 \\ \noalign{\smallskip}
    20    & 4p$^{5}$5p $(^{1}$P$_{1}) $          & 1.99657 & 2.07309 & \phantom{-}3.691 \\ \noalign{\smallskip}
    21    & 4p$^{5}$5p $(^{3}$P$_{2}) $          & 2.00578 & 2.08260 & \phantom{-}3.689 \\ \noalign{\smallskip}
    22    & 4p$^{5}$4d $(^{1}$P$_{1}^{\circ})$   & 2.08800 & 2.08415 & -0.185 \\ \noalign{\smallskip}
    23    & 4p$^{5}$5p $(^{3}$P$_{0}) $          & 2.05042 & 2.11630 & \phantom{-}3.113 \\ \noalign{\smallskip}
    24    & 4p$^{5}$5p $(^{3}$D$_{1}) $          & 2.06224 & 2.13923 & \phantom{-}3.599 \\ \noalign{\smallskip}
    25    & 4p$^{5}$5p $(^{3}$D$_{2}) $          & 2.07817 & 2.15639 & \phantom{-}3.627 \\ \noalign{\smallskip}
    26    & 4p$^{5}$5p $(^3$P$_{1}) $            & 2.08289 & 2.15839 & \phantom{-}3.498 \\ \noalign{\smallskip}
    27    & 4p$^{5}$5p $(^{1}$S$_{0}) $          & 2.16041 & 2.19616 & \phantom{-}1.628 \\ \noalign{\smallskip}
    28    & 4p$^{5}$5d $(^3$P$_{0}^{\circ}) $    & 2.36254 & 2.44628 & \phantom{-}3.424 \\ \noalign{\smallskip}
    29    & 4p$^{5}$6s $(^3$P$_{2}^{\circ}) $    & 2.36912 & 2.44821 & \phantom{-}3.230 \\ \noalign{\smallskip}
    30    & 4p$^{5}$5d $(^3$P$_{1}^{\circ}) $    & 2.36869 & 2.45095 & \phantom{-}3.357 \\ \noalign{\smallskip}
    \hline
    \noalign{\smallskip}
    \end{tabular}%
    \label{tab:SrIII-Energy-Levels}%
\end{table}

\begin{figure}
\centering

\includegraphics[width=1.0\columnwidth]{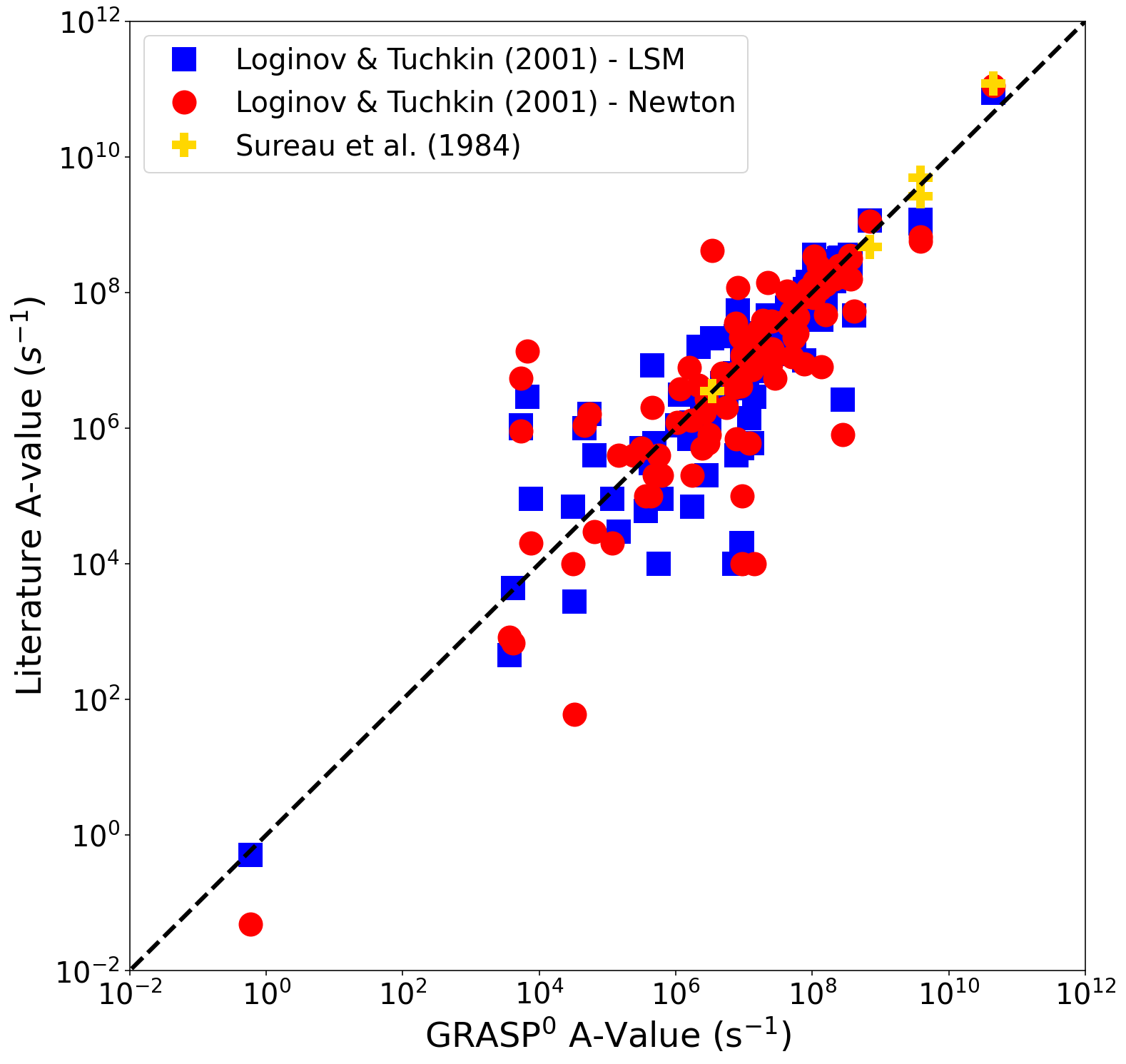}
\caption{A comparison of the Einstein A-values for transitions from the {\sc grasp$^0$} Sr {\sc iii} model and their equivalents from other calculations in the literature, \protect\cite{sureau_1984} and \protect\cite{Loginov_2001}.}
\label{fig:SrIII_A-values}.
\end{figure}

A search of the literature found no A-values determined through experimental means, so we instead make a comparison with other theoretical calculations. \cite{Loginov_2001} performed two separate calculations, a least-squares method and an application of Newton's method. In addition, five Sr {\sc iii} A-values were cited in \cite{sureau_1984}, which employed a theoretical Self-Consistent Field (SCF) approach in a multi-configuration Hartee-Fock (MCHF) calculation. The comparison between the A-values from those computed in this work and the three literature sources is highlighted in \text{Figure \ref{fig:SrIII_A-values}} where again we see good agreement along the line of equality, particularly for the strongest lines. Excellent agreement is evident with the five individual lines considered in the MCHF approach of \cite{sureau_1984}. We also present specific examples in  \text{Table \ref{tab:SrIII-A-values}} corresponding to the transitions determined in \cite{sureau_1984} and overall there is good agreement found between the calculations. The \cite{sureau_1984} data especially shows very strong agreement, whereas both data sets from \cite{Loginov_2001} exhibit slightly higher degrees of variation for individual transitions. The majority of the largest differences correspond to the weaker transitions. 

\begin{table}
  \centering
  \caption{Comparison between sample Einstein A-values obtained from the Sr {\sc iii} GRASP$^0$ target and the literature ([1]- \protect\cite{sureau_1984} and [2]- \protect\cite{Loginov_2001} - LSM Method).  The Index column refers to the energy levels displayed in \text{Table \ref{tab:SrIII-Energy-Levels}}. The wavelengths corresponding to each transition were experimentally determined in \protect\cite{Reader_1972}.}
    \begin{tabular}{cccc} \hline
            \textbf{Index} & \TableNewLine{\textbf{Wavelength} \\ \textbf{ / nm}}      & \TableNewLine{\textbf{GRASP$^0$} \\ \textbf{A-value / s$^{-1}$} } & \TableNewLine{\textbf{Literature} \\ \textbf{A-value / s$^{-1}$}}  \\ \hline
 \noalign{\smallskip}
    1 - 3\phantom{1}     & 56.2752  & 3.38E+06 & 3.50E+06$^{[1]}$ \\ \noalign{\smallskip}
    & & & 2.09E+07$^{[2]}$ \\ \noalign{\smallskip}
    1 - 10               & 51.4376  & 3.87E+09 & 2.60E+09$^{[1]}$ \\ \noalign{\smallskip}
    & & & 1.17E+09$^{[2]}$ \\ \noalign{\smallskip}
    1 - 12               & 50.7035  & 6.90E+08 & 4.70E+08$^{[1]}$ \\ \noalign{\smallskip}
    & & & 1.14E+09$^{[2]}$ \\ \noalign{\smallskip}
    1 - 16               & 49.1786  & 3.88E+09 & 4.90E+09$^{[1]}$ \\ \noalign{\smallskip}
    & & & 1.05E+09$^{[2]}$ \\ \noalign{\smallskip}
    1 - 22               & 43.7240  & 4.46E+10 & 1.20E+11$^{[1]}$ \\ \noalign{\smallskip}
    & & & 8.68E+10$^{[2]}$ \\ \noalign{\smallskip}
    \hline
    \noalign{\smallskip}
    \end{tabular}%
    \label{tab:SrIII-A-values}%
\end{table}

\subsection{Sr {\sc iv} Model} \label{sec:SrIV_Structure}

Sr {\sc iv} is part of the Bromine isoelectronic sequence, and as such consists of 35 electrons.  It has a ground state configuration of [Ar]3d$^{10}$4s$^2$4p$^5$ ($^2$P$_{3/2}^{\circ}$).  The {\sc grasp$^0$} model developed for Sr {\sc iv} consists of 19 orbitals extending to $n$=7 and $l$=3 (1s, 2s, 2p, 3s, 3p, 3d, 4s, 4p, 4d, 4f, 5s, 5p, 5d, 5f, 6s, 6p, 6d, 7s and 7p).  The target consists of the 21 configurations listed in \text{Table \ref{tab:Sr_IV_Structure}} and includes not only single electron promotions out the 4p$^5$ ground state but also from the 4s$^2$ inner orbital. These configurations resulted in a target comprising a total of 504 fine-structure levels. 

NIST provides experimental data for 255 individual energy levels for Sr {\sc iv} from the works of \cite{Persson_1978} and \cite{sansonetti_2012}.  We matched the 70 lowest lying levels spanning the energy range from 0.0 - 3.0 Ryd, exceeding the temperatures observed in KNe events.  \text{Table \ref{tab:Sr_IV_Energy_Levels}} displays the comparison between the lowest 30 of these energy levels and shows good agreement between the current values and their equivalents in NIST. The differences are primarily $<$0.01 Ryd, although for levels 23, 24 and 30 larger disparities are evident. The average relative percentage difference across all 70 shifted energy levels is -0.720\%.

   \begin{table}
    \centering
    \caption[]{The configurations included in the wavefunction expansion for the GRASP$^0$ model for Sr {\sc iv}.}
         \begin{tabular}{p{2.0cm}p{2.0cm}p{2.0cm}}
            \hline
            \noalign{\smallskip}
            \multicolumn{3}{l}{\textbf{Sr {\sc iv} - 21 Configurations}} \\
            \noalign{\smallskip}
            \hline
            \noalign{\smallskip}
             4s$^{2}$4p$^{5}$     & 4s$^{2}$4p$^{4}$4d       & 4s$^{2}$4p$^{4}$4f        \\ \noalign{\smallskip}
             4s$^{2}$4p$^{4}$5s   & 4s$^{2}$4p$^{4}$5p       & 4s$^{2}$4p$^{4}$5d        \\ \noalign{\smallskip}
             4s$^{2}$4p$^{4}$5f   & 4s$^{2}$4p$^{4}$6s       & 4s$^{2}$4p$^{4}$6p        \\ \noalign{\smallskip}
             4s$^{2}$4p$^{4}$6d   & 4s$^{2}$4p$^{4}$7s       & 4s$^{2}$4p$^{4}$7p        \\ \noalign{\smallskip}
             4s4p$^{6}$           & 4s4p$^{5}$4d             & 4s4p$^{5}$4f              \\ \noalign{\smallskip}
             4s4p$^{5}$5s         & 4s4p$^{5}$5p             & 4p$^{5}$6d$^{2}$          \\ \noalign{\smallskip}
             4s4p$^{4}$5s6s       & 4s$^{2}$4p$^{3}$5p$^{2}$ & 4s$^{2}$4p$^{3}$6p$^{2}$  \\ \noalign{\smallskip}
            \hline
            \noalign{\smallskip}
         \end{tabular}
        \label{tab:Sr_IV_Structure}
   \end{table}

\begin{table}
\centering
\caption[]{The first 30 energy levels of the Sr {\sc iv} model.  The NIST energies for the first 30 energy levels were obtained from \protect\cite{Persson_1978}.}
%\resizebox{2\columnwidth}{!}{%
\begin{tabular}{ccccc}\hline
\textbf{Level} & \textbf{Config.}      & \TableNewLine{\textbf{GRASP$^0$} \\ \textbf{/ Ryd}} & \TableNewLine{\textbf{NIST} \\ \textbf{/ Ryd}} & \TableNewLine{\textbf{Percentage} \\ \textbf{Error / \%}} \\\hline
\noalign{\smallskip}
1     &    4s$^{2}$4p$^{5}$ $(^2$P$_{3/2}^{\circ})$     & 0.00000    &  0.00000  &                    \\  \noalign{\smallskip}
2     &    4s$^{2}$4p$^{5}$ $(^2$P$_{1/2}^{\circ})$     & 0.08854    &  0.08865  & \phantom{-}0.120   \\ \noalign{\smallskip}
3     &    4s4p$^{6}$     $(^2$S$_{1/2})$               & 1.37586    &  1.37149  & -0.318   \\ \noalign{\smallskip}
4     &    4p$^{4}$4d $(^4$D$_{7/2})$                   & 1.70536    &  1.70553  & \phantom{-}0.010   \\ \noalign{\smallskip}
5     &    4p$^{4}$4d $(^4$D$_{5/2})$                   & 1.70697    &  1.70590  & -0.063              \\ \noalign{\smallskip}
6     &    4p$^{4}$4d $(^4$D$_{3/2})$                   & 1.71527    &  1.71340  & -0.109   \\ \noalign{\smallskip}
7     &    4p$^{4}$4d $(^4$D$_{1/2})$                   & 1.72559    &  1.72339  & -0.128   \\ \noalign{\smallskip}
8     &    4p$^{4}$4d $(^4$F$_{9/2})$                   & 1.80900    &  1.79574  & -0.738              \\ \noalign{\smallskip}
9     &    4p$^{4}$4d $(^4$F$_{7/2})$                   & 1.84166    &  1.82563  & -0.878              \\ \noalign{\smallskip}
10    &    4p$^{4}$4d $(^2$P$_{1/2})$                   & 1.86568    &  1.82736  & -2.097   \\ \noalign{\smallskip}
11    &    4p$^{4}$4d $(^4$F$_{5/2})$                   & 1.86683    &  1.85301  & -0.746              \\ \noalign{\smallskip}
12    &    4p$^{4}$4d $(^4$F$_{3/2})$                   & 1.87741    &  1.86062  & -0.902              \\ \noalign{\smallskip}
13    &    4p$^{4}$4d $(^4$P$_{1/2})$                   & 1.89299    &  1.86576  & -1.460   \\ \noalign{\smallskip}
14    &    4p$^{4}$4d $(^4$P$_{3/2})$                   & 1.89677    &  1.86636  & -1.630   \\ \noalign{\smallskip}
15    &    4p$^{4}$4d $(^2$D$_{3/2})$                   & 1.91782    &  1.88199  & -1.904   \\ \noalign{\smallskip}
16    &    4p$^{4}$4d $(^2$F$_{7/2})$                   & 1.92479    &  1.89068  & -1.804   \\ \noalign{\smallskip}
17    &    4p$^{4}$4d $(^4$P$_{5/2})$                   & 1.92975    &  1.90398  & -1.353   \\ \noalign{\smallskip}
18    &    4p$^{4}$4d $(^2$P$_{3/2})$                   & 1.94113    &  1.90647  & -1.818              \\ \noalign{\smallskip}
19    &    4p$^{4}$4d $(^2$D$_{5/2})$                   & 1.96658    &  1.93164  & -1.809   \\ \noalign{\smallskip}
20    &    4p$^{4}$4d $(^2$F$_{5/2})$                   & 1.99858    &  1.95873  & -2.035              \\ \noalign{\smallskip}
21    &    4p$^{4}$4d $(^2$G$_{9/2})$                   & 2.00131    &  1.96093  & -2.059   \\ \noalign{\smallskip}
22    &    4p$^{4}$4d $(^2$G$_{7/2})$                   & 2.00368    &  1.96094  & -2.180   \\ \noalign{\smallskip}
23    &    4p$^{4}$4d $(^2$F$_{5/2})$                   & 2.12542    &  2.05829  & -3.261   \\ \noalign{\smallskip}
24    &    4p$^{4}$4d $(^2$F$_{7/2})$                   & 2.14392    &  2.07858  & -3.143              \\ \noalign{\smallskip}
25    &    4p$^{4}$5s $(^4$P$_{5/2})$                   & 2.07754    &  2.08365  & \phantom{-}0.293              \\ \noalign{\smallskip}
26    &    4p$^{4}$5s $(^4$P$_{3/2})$                   & 2.11342    &  2.11606  & \phantom{-}0.125              \\ \noalign{\smallskip}
27    &    4p$^{4}$5s $(^4$P$_{1/2})$                   & 2.15364    &  2.15902  & \phantom{-}0.249              \\ \noalign{\smallskip}
28    &    4p$^{4}$5s $(^2$P$_{3/2})$                   & 2.16874    &  2.17080  & \phantom{-}0.095              \\ \noalign{\smallskip}
29    &    4p$^{4}$5s $(^2$P$_{1/2})$                   & 2.21043    &  2.20611  & -0.196              \\ \noalign{\smallskip}
30    &    4p$^{4}$4d $(^2$D$_{3/2})$                   & 2.30487    &  2.20683  & -4.443             \\ \noalign{\smallskip}
\hline
\noalign{\smallskip}
\end{tabular}%
\label{tab:Sr_IV_Energy_Levels}
%}
\end{table}

\begin{figure}
\centering

\includegraphics[width=1.0\columnwidth]{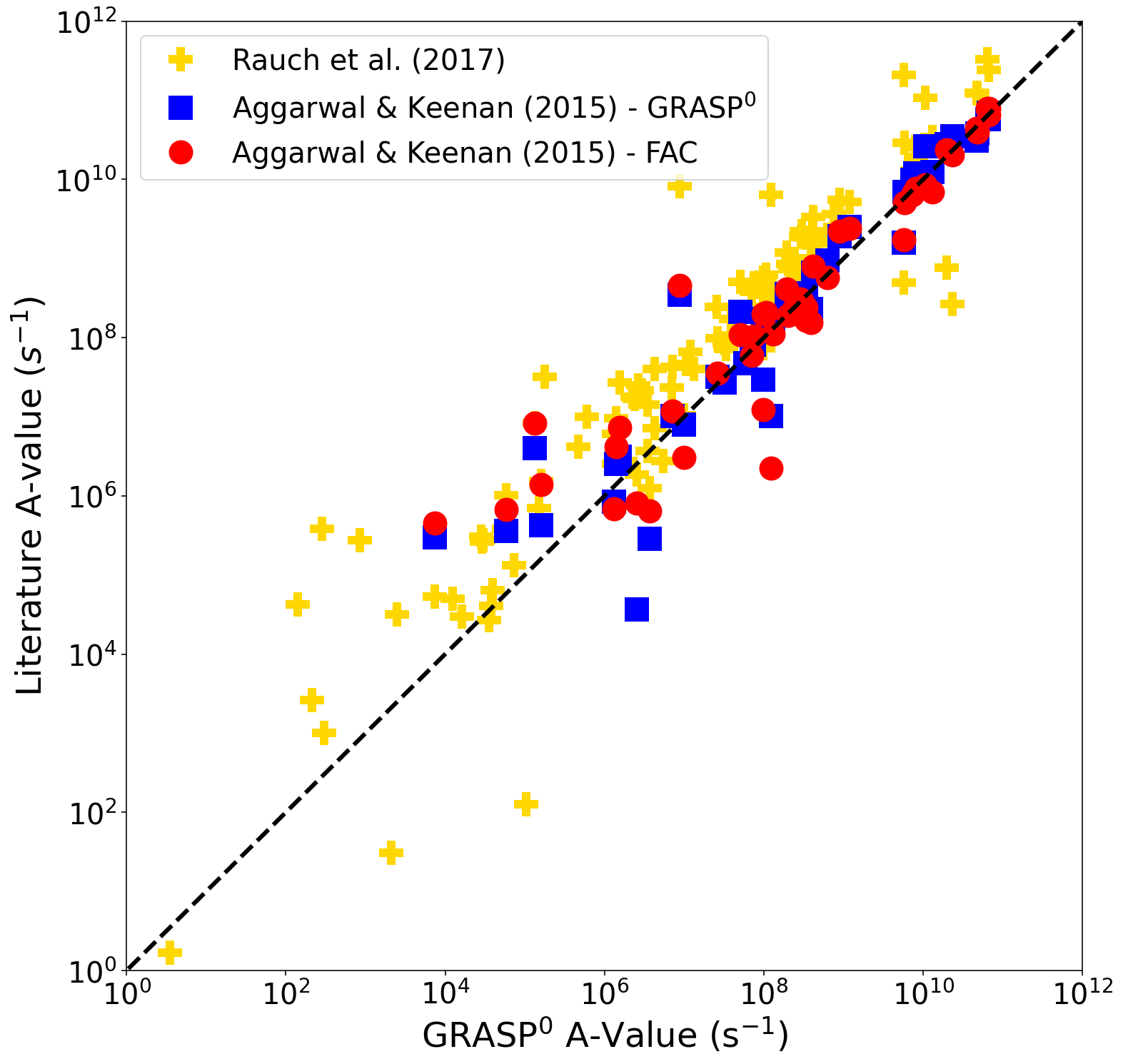}
\caption{A comparison of the Einstein A-values for transitions from the GRASP$^0$ Sr {\sc iv} model and the literature A-values of \protect\cite{Aggarwal_2015} and \protect\cite{Rauch_2017}.}
\label{fig:SrIV_A-values}.
\end{figure}

In \text{Figure \ref{fig:SrIV_A-values}} we compare the current A-values with those calculated in \cite{Aggarwal_2015}, who also employed a {\sc grasp$^0$} methodology using a different target Sr {\sc iv} target structure, with unshifted energy levels.  \cite{Aggarwal_2015} compared their results to those from a second Sr {\sc iv} model constructed using the Flexible Atomic Code (FAC, \cite{Gu_2008}). In \text{Figure \ref{fig:SrIV_A-values}} we also compare the A-values determined in \cite{Rauch_2017}, which were calculated using a modification of the Cowan Code (\cite{Cowan_1981}, \cite{Quinet_1999}). This implementation applies a semi-relativistic Hartree-Fock method with core polarisation effects.  The comparison for selected transitions with A-values between 10$^{0}$ and 10$^{12}$ s$^{-1}$, shows good agreement between all three calculations and the current values. A selection of strong dipole lines are presented in \text{Table \ref{tab:SrIV-A-values}} where good agreement is evident with the GRASP$^0$ values of \cite{Aggarwal_2015}. The differences in select A-values highlights the significance of shifting the energy levels to their spectroscopic position prior to the evaluation of the A-values. Agreement with the predictions of \cite{Rauch_2017} is less satisfactory, their A-values being systematically larger than those determined in this work and in the evaluations of \cite{Aggarwal_2015}.

\begin{table}

  \centering
   \caption{Comparison of sample Einstein A-values obtained from the Sr {\sc iv} GRASP$^0$ target and those obtained from [1]-\protect\cite{Aggarwal_2015} and [2]-\protect\cite{Rauch_2017}.  The Index column refers to the energy levels displayed in \text{Table \ref{tab:Sr_IV_Energy_Levels}}. The wavelengths corresponding to each transition were experimentally measured in \protect\cite{Hansen_1976}. }
    \begin{tabular}{cccc} \hline
            \textbf{Index} & \TableNewLine{\textbf{Wavelength} \\ \textbf{ / nm}}      & \TableNewLine{\textbf{AS} \\ \textbf{A-value / s$^{-1}$} } & \TableNewLine{\textbf{Literature}\\ \textbf{A-value / s$^{-1}$} }  \\ \hline
 \noalign{\smallskip}
    1 - 3\phantom{-}     & 66.4434 & 1.92E+08 & 3.55E+08$^{[1]}$ \\ \noalign{\smallskip}
         &  &  & 1.19E+09$^{[2]}$ \\ \noalign{\smallskip}
    1 - 14     & 48.8261 & 1.92E+08 & 1.06E+08$^{[1]}$ \\ \noalign{\smallskip}
         &  &  & 6.19E+08$^{[2]}$ \\ \noalign{\smallskip}
    1 - 19     & 47.1758 & 2.30E+08 & 3.89E+08$^{[1]}$ \\ \noalign{\smallskip}
         &  &  & 9.97E+08$^{[2]}$ \\ \noalign{\smallskip}
    1 - 26     & 43.0645 & 5.82E+09 & 6.90E+09$^{[1]}$ \\ \noalign{\smallskip}
        &  &  & 2.90E+10$^{[2]}$ \\ \noalign{\smallskip}
    2 - 15     & 50.8140 & 2.36E+08 & 2.53E+08$^{[1]}$ \\ \noalign{\smallskip}
         &  &  & 4.98E+08$^{[2]}$ \\ \noalign{\smallskip}
    2 - 26     & 44.9478 & 2.01E+08 & 2.64E+08$^{[1]}$ \\ \noalign{\smallskip}
         &  &  & 8.48E+08$^{[2]}$ \\ \noalign{\smallskip}
    \hline
    \noalign{\smallskip}
    \end{tabular}%
    \label{tab:SrIV-A-values}%
\end{table}

\subsection{Sr {\sc v} Model} \label{sec:Sr_V_Structure}

Sr {\sc v} is a part of the Selenium isoelectronic series, and as such consists of 34 electrons.  It has a ground state configuration of [Ar] 3d$^{10}$4s$^2$4p$^4$ ($^3$P$_2$). The AS model for Sr {\sc v} consists of 24 orbitals up to $n$=7 and $l$=3 (1s, 2s, 2p, 3s, 3p, 3d, 4s, 4p, 4d, 4f, 5s, 5p, 5d, 5f, 6s, 6p, 6d, 6f, 7s and 7p). The target model, comprising 23 configurations as listed in \text{Table \ref{tab:Sr_V_Structure}}, includes single and double promotions from the 4s and 4p subshells. This target model gave rise to a substantial 1817 distinct fine-structure levels and all the $\lambda_{nl}$ parameters adopted in the AS computations were set to a value of 1.0, with the exception of $\lambda_{5f}$ which was assigned a value of 1.050. These scaling parameter values were found to produce the best energy levels for the target states included.

NIST contained 144 energy levels for Sr {\sc v}, obtained from the works of \cite{Persson_1984} and \cite{sansonetti_2012}. We identified and matched the 95 lowest lying energy levels and \text{Table \ref{tab:Sr_V_Energy_Levels}} displays the comparison for the first 30 of these.  There is very good agreement between the {\sc grasp}$^{0}$ predictions and the NIST values, with the average relative percentage difference for all 95 shifted energy levels being 1.25\%.

   \begin{table}
    \centering
    \caption[]{The configurations included in the wavefunction expansion for the {\sc grasp$^0$} model for Sr {\sc v}.}
         \begin{tabular}{p{2.0cm}p{2.0cm}p{2.0cm}}
            \hline
            \noalign{\smallskip}
            \multicolumn{3}{l}{\textbf{Sr {\sc v} - 23 Configurations}} \\
            \noalign{\smallskip}
            \hline
            \noalign{\smallskip}
             4s$^{2}$4p$^{4}$         &  4s4p$^{5}$              & 4s$^{2}$4p$^{3}$4d       \\ \noalign{\smallskip}
             4s$^{2}$4p$^{3}$4f       & 4s$^{2}$4p$^{3}$5s       & 4s$^{2}$4p$^{3}$5p       \\ \noalign{\smallskip}
             4s$^{2}$4p$^{3}$5d       & 4s$^{2}$4p$^{3}$5f       & 4s$^{2}$4p$^{3}$6s       \\ \noalign{\smallskip}
             4s$^{2}$4p$^{3}$6p       & 4s$^{2}$4p$^{3}$6d       & 4s$^{2}$4p$^{3}$6f       \\ \noalign{\smallskip}
             4s$^{2}$4p$^{3}$7s       & 4s$^{2}$4p$^{3}$7p       & 4s$^{2}$4p$^{2}$4d4f     \\ \noalign{\smallskip}
             4s$^{2}$4p$^{2}$4d5s     & 4s$^{2}$4p$^{2}$4d5p     & 4s$^{2}$4p$^{2}$4d5d     \\ \noalign{\smallskip}
             4s$^{2}$4p$^{2}$4d5f     & 4s$^{2}$4p$^{2}$4d6s     & 4s$^{2}$4p4d$^{3}$       \\ \noalign{\smallskip}
             4p$^{4}$4d4f             & 4p$^{5}$4d               & {}                       \\ \noalign{\smallskip}
            \hline
            \noalign{\smallskip}
         \end{tabular}
        
        \label{tab:Sr_V_Structure}
   \end{table}

\begin{table}
\centering
\caption[]{The first 30 energy levels of the Sr {\sc v} model.  The average percentage between the shifted and unshifted levels was 1.250\%.  The NIST energies for the first 30 energy levels were obtained from \protect\cite{Persson_1984}.}
%\resizebox{2\columnwidth}{!}{%
\begin{tabular}{ccccc}\hline
\textbf{Level} & \textbf{Config.}      & \TableNewLine{\textbf{AS} \\ \textbf{/ Ryd}} & \TableNewLine{\textbf{NIST} \\ \textbf{/ Ryd}} & \TableNewLine{\textbf{Percentage} \\ \textbf{Error / \%}} \\\hline
\noalign{\smallskip}
1     &                4s$^{2}$4p$^{4}$ $(^3$P$_2)$     & 0.00000    &  0.00000  &            \\  \noalign{\smallskip}
2     &                4s$^{2}$4p$^{4}$ $(^3$P$_1)$     & 0.09422    &  0.07571  & \phantom{-}24.448  \\ \noalign{\smallskip}
3     &                4s$^{2}$4p$^{4}$ $(^3$P$_0)$     & 0.09688    &  0.07945  & \phantom{-}21.940  \\ \noalign{\smallskip}
4     &                4s$^{2}$4p$^{4}$ $(^1$D$_0)$     & 0.22358    &  0.18508  & \phantom{-}20.801  \\ \noalign{\smallskip}
5     &                4s$^{2}$4p$^{4}$ $(^1$S$_2)$     & 0.48129    &  0.40142  & \phantom{-}19.898  \\ \noalign{\smallskip}
6     &            4s4p$^{5}$ $(^3$P$_{2}^{\circ})$     & 1.34914    &  1.40365  & -3.883   \\ \noalign{\smallskip}
7     &            4s4p$^{5}$ $(^3$P$_{1}^{\circ})$     & 1.41684    &  1.45819  & -2.836   \\ \noalign{\smallskip}
8     &            4s4p$^{5}$ $(^3$P$_{0}^{\circ})$     & 1.46288    &  1.49462  & -2.124   \\ \noalign{\smallskip}
9     &            4s4p$^{5}$ $(^1$P$_{1}^{\circ})$     & 1.74776    &  1.76165  & -0.789   \\ \noalign{\smallskip}
10    &    4s$^{2}$4p$^{3}$4d $(^5$D$_{0}^{\circ})$     & 1.79679    &  1.84439  & -2.581   \\ \noalign{\smallskip}
11    &    4s$^{2}$4p$^{3}$4d $(^5$D$_{1}^{\circ})$     & 1.79702    &  1.84193  & -2.438   \\ \noalign{\smallskip}
12    &    4s$^{2}$4p$^{3}$4d $(^5$D$_{2}^{\circ})$     & 1.79764    &  1.84369  & -2.498   \\ \noalign{\smallskip}
13    &    4s$^{2}$4p$^{3}$4d $(^5$D$_{3}^{\circ})$     & 1.79824    &  1.84318  & -2.438   \\ \noalign{\smallskip}
14    &    4s$^{2}$4p$^{3}$4d $(^5$D$_{4}^{\circ})$     & 1.80058    &  1.84908  & -2.623   \\ \noalign{\smallskip}
15    &    4s$^{2}$4p$^{3}$4d $(^3$D$_{2}^{\circ})$     & 1.92038    &  1.94814  & -1.425   \\ \noalign{\smallskip}
16    &    4s$^{2}$4p$^{3}$4d $(^3$D$_{3}^{\circ})$     & 1.95266    &  1.96929  & -0.844   \\ \noalign{\smallskip}
17    &    4s$^{2}$4p$^{3}$4d $(^3$D$_{1}^{\circ})$     & 1.96627    &  1.97717  & -0.551   \\ \noalign{\smallskip}
18    &    4s$^{2}$4p$^{3}$4d $(^3$F$_{2}^{\circ})$     & 1.99773    &  2.00980  & -0.600   \\ \noalign{\smallskip}
19    &    4s$^{2}$4p$^{3}$4d $(^1$S$_{0}^{\circ})$     & 2.01609    &  2.02637  & -0.507   \\ \noalign{\smallskip}
20    &    4s$^{2}$4p$^{3}$4d $(^3$F$_{3}^{\circ})$     & 2.03695    &  2.02099  & \phantom{-}0.789   \\ \noalign{\smallskip}
21    &    4s$^{2}$4p$^{3}$4d $(^3$F$_{4}^{\circ})$     & 2.03947    &  2.04893  & -0.462   \\ \noalign{\smallskip}
22    &    4s$^{2}$4p$^{3}$4d $(^3$G$_{3}^{\circ})$     & 2.09020    &  2.09545  & -0.251   \\ \noalign{\smallskip}
23    &    4s$^{2}$4p$^{3}$4d $(^3$G$_{4}^{\circ})$     & 2.10168    &  2.10479  & -0.148   \\ \noalign{\smallskip}
24    &    4s$^{2}$4p$^{3}$4d $(^3$G$_{5}^{\circ})$     & 2.12157    &  2.11948  & \phantom{-}0.099   \\ \noalign{\smallskip}
25    &    4s$^{2}$4p$^{3}$4d $(^1$G$_{4}^{\circ})$     & 2.14721    &  2.13950  & \phantom{-}0.360   \\ \noalign{\smallskip}
26    &    4s$^{2}$4p$^{3}$4d $(^1$D$_{2}^{\circ})$     & 2.17734    &  2.20052  & -1.054   \\ \noalign{\smallskip}
27    &    4s$^{2}$4p$^{3}$4d $(^3$D$_{1}^{\circ})$     & 2.19996    &  2.21963  & -0.886    \\ \noalign{\smallskip}
28    &    4s$^{2}$4p$^{3}$4d $(^3$P$_{0}^{\circ})$     & 2.23367    &  2.24902  & -0.683    \\ \noalign{\smallskip}
29    &    4s$^{2}$4p$^{3}$4d $(^3$P$_{1}^{\circ})$     & 2.24012    &  2.26005  & -0.882    \\ \noalign{\smallskip}
30    &    4s$^{2}$4p$^{3}$4d $(^3$D$_{2}^{\circ})$     & 2.24712    &  2.26344  & -0.721   \\ \noalign{\smallskip}
\hline
\noalign{\smallskip}
\end{tabular}%
\label{tab:Sr_V_Energy_Levels}
\end{table}

\begin{figure}
\centering

\includegraphics[width=1.0\columnwidth]{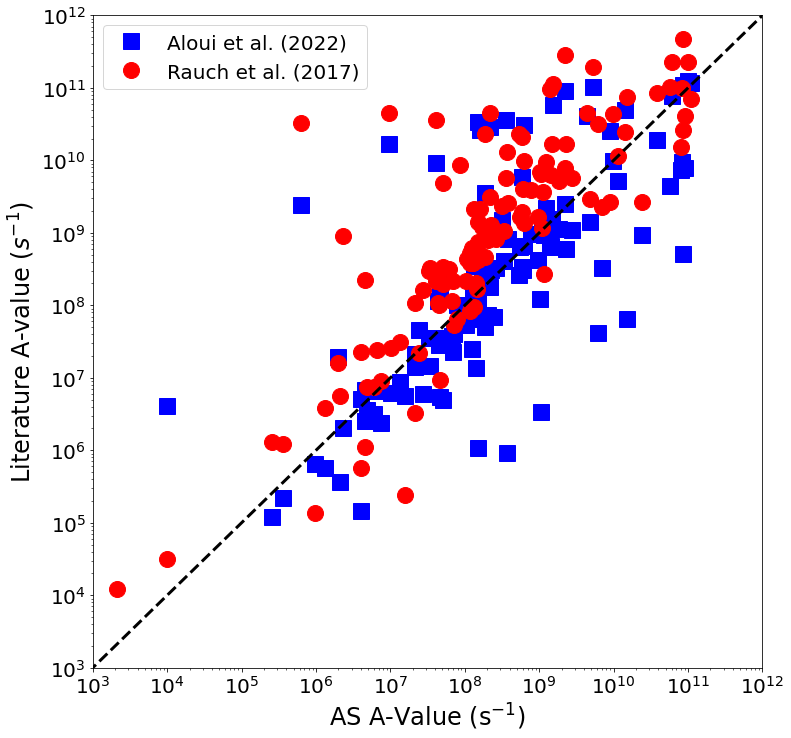}
\caption{A comparison of the Einstein A-values for dipole transitions from the AS Sr {\sc v} model and their equivalents from other structure calculations in the literature.  The literature A-values were obtained from \protect\cite{Rauch_2017}. and \protect\cite{Aloui_2022}.}
\label{fig:SrV_A-values}
\end{figure}

Theoretical A-values for Sr {\sc v} were determined in \cite{Rauch_2017} using a modification of the Cowan code and by \cite{Aloui_2022} who utilised the {\sc superstructure} (\cite{eissner_1991}) package for use in determining Stark widths of Sr {\sc v} lines in the atmospheres of hot WDs.  The latter calculations consisted of five configurations, all of which are included in the present Sr {\sc v} target. In \text{Figure \ref{fig:SrV_A-values}} we compare A-values for the dipole transitions among the first 50 levels to those found in the two data sets from the literature. Good agreement is found, particularly with the predictions of \cite{Aloui_2022} but there are some notable differences for some weaker transitions. These disparities may result from the different Sr {\sc v} targets adopted and/or errors arising between shifted and non-shifted energy levels prior to the computation of the A-values. Reasonable agreement is also found with the \cite{Rauch_2017} data set, but as with Sr {\sc iv}, the AS values tend to be of lower magnitude compared to their equivalents from the Cowan runs. In \text{Table \ref{tab:SrV-A-values}} a selection of the strongest lines are displayed which have been identified by \cite{Rauch_2017} in the spectra of the hot WD star RE 0503-289. In general, good agreement is found for those transitions where a comparison is possible and particularly between the present {\sc grasp}$^{0}$ values and those computed by \cite{Aloui_2022}.

\begin{table}

  \centering
      \caption{Comparison between sample Einstein A-values obtained from the Sr {\sc v} AS target and those obtained from [1]-\protect\cite{Rauch_2017} and [2]-\protect\cite{Aloui_2022}.  The Index column refers to the energy levels displayed in Table \ref{tab:Sr_V_Energy_Levels}. The wavelengths corresponding to each transition were experimentally measured in \protect\cite{Persson_1984} and identified in \protect\cite{Rauch_2017}. }
    \begin{tabular}{cccc} \hline
            \textbf{Index} & \TableNewLine{\textbf{Wavelength} \\ \textbf{ / nm}}      & \TableNewLine{\textbf{AS} \\ \textbf{A-value / s$^{-1}$} } & \TableNewLine{\textbf{Literature}\\ \textbf{A-value / s$^{-1}$} }  \\ \hline
 \noalign{\smallskip}
    1 - 29     & 40.3204 & 2.73E+09 & 7.88E+09$^{[1]}$ \\ \noalign{\smallskip}
         &  &  & 1.11E+09$^{[2]}$ \\ \noalign{\smallskip}
    3 - 27     & 57.8006 & 2.22E+09 & 7.88E+09$^{[1]}$ \\ \noalign{\smallskip}
        &  &  & 2.47E+09$^{[2]}$ \\ \noalign{\smallskip}
    4 - 9\phantom{-}     & 42.5790 & 1.14E+09 & 3.68E+09$^{[1]}$ \\ \noalign{\smallskip}
         &  &  & 1.51E+09$^{[2]}$ \\ \noalign{\smallskip}
    33 - 81    & 92.2397 & 2.181E+09 & 1.25E+10$^{[1]}$ \\ \noalign{\smallskip}
         &  &  & 2.68E+09$^{[2]}$ \\ \noalign{\smallskip}
    34 - 70    & 115.4871 & 1.939E+08 & 1.12E+09$^{[1]}$ \\ \noalign{\smallskip}
         &  &  & 1.07E+08$^{[2]}$ \\ \noalign{\smallskip}
    37 - 84     & 101.3714 & 9.172E+08 & 3.49E+09$^{[1]}$ \\ \noalign{\smallskip}
        &  &  & 3.07E+06$^{[2]}$ \\ \noalign{\smallskip}
    57 - 90     & 137.2843 & 1.961E+09 & 5.79E+09$^{[1]}$ \\ \noalign{\smallskip}
    60 - 91     & 138.0477 & 8.306E+08 & 7.09E+09$^{[1]}$ \\ \noalign{\smallskip}
    60 - 95     & 141.3882 & 2.527E+09 & 2.20E+10$^{[1]}$ \\ \noalign{\smallskip}
    \hline
    \noalign{\smallskip}
    \end{tabular}%
    \label{tab:SrV-A-values}%
\end{table}

\begin{figure*}
\centering
\includegraphics[scale=0.40]{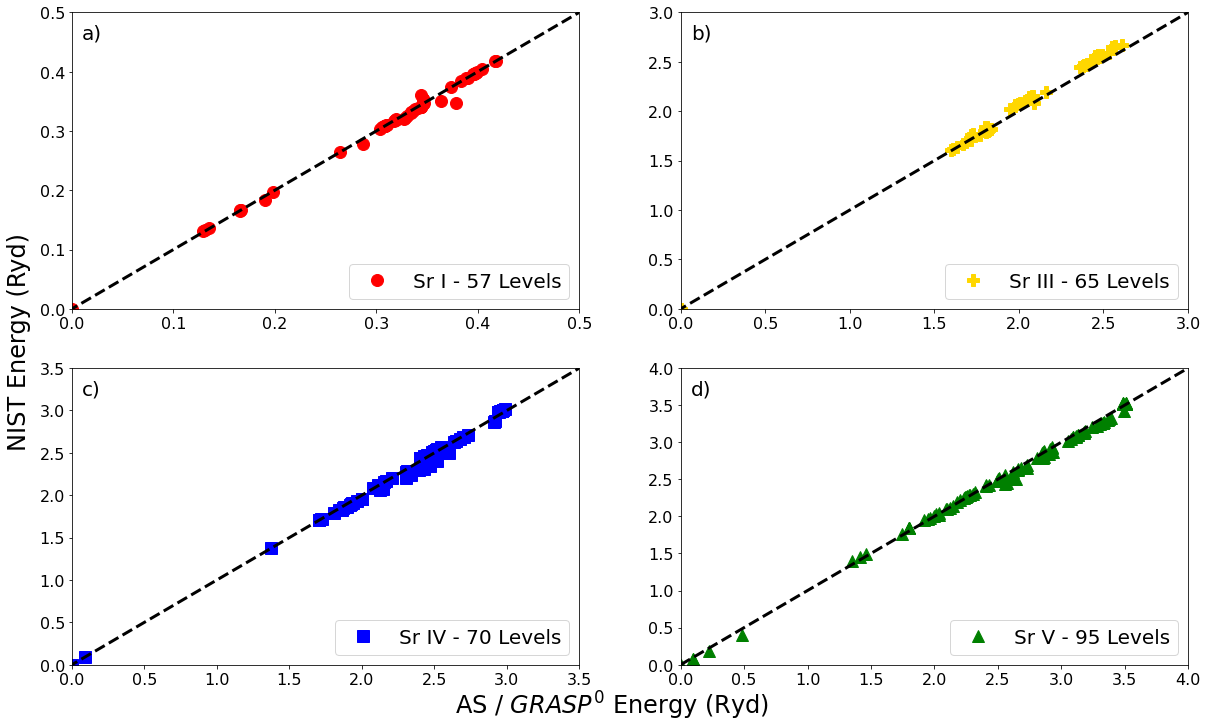}
\caption{Comparison between the energy levels derived from the AS and {\sc grasp$^0$} models of a) - Sr {\sc i}, b) - Sr {\sc iii}, c) - Sr {\sc iv} and d) - Sr {\sc v}, compared to their equivalent values obtained from NIST.}
\label{fig:Sr_Energy_Comparison}
\end{figure*}

In \text{Figure \ref{fig:SrI_A-values}}, \text{Figure \ref{fig:SrIII_A-values}}, \text{Figure \ref{fig:SrIV_A-values}} and \text{Figure \ref{fig:SrV_A-values}} we show graphically the conformity between the A-values computed in this work with all available data currently available in the literature. For completeness, we present in \text{Figure \ref{fig:Sr_Energy_Comparison}} similar graphical evidence for the accuracy of the energy levels for each Sr species considered. We note that the collision calculations discussed in the next section were computed with the energy levels calibrated to their spectroscopic positions where possible. This will ensure a more accurate identification of lines during the analysis of observational spectra.

%In the following sections, we will discuss each of the four strontium models.  This will include which orbitals and configurations were included in the wavefunction expansion of the target.  A comparison between the energy levels and A-values obtained from each target is made with other literature values as a test of validity.  \textit{Section \ref{sec:SrI_Structure}} will describe the AS Sr I target, \textit{Section \ref{sec:SrIII_Model}} the GRASP$^0$ target of Sr III, \textit{Section \ref{sec:SrIV_Structure}} the GRASP$^0$ target of Sr IV and \textit{Section \ref{sec:Sr_V_Structure}} the AS model of Sr V. 

\section{Electron Impact Excitation}
\label{sec:Electron-Impact-Excitation}
\subsection{$R$-matrix Methodology}

The $R$-matrix approach is employed for the collisional calculations to determine the electron-impact excitation rates for each of the transitions in the Sr models discussed above. A brief overview of the theory involved is outlined here, but a full description of the procedure may be found in \cite{Burke_2011}.

In the $R$-matrix method, configuration space describing the atomic system is divided into two regions, internal and external, with the $N$-electron target positioned at the centre.  The incident and subsequent scattered electron is represented by a different wavefunction in each region, representations that remain continuous across the boundary.  In the inner region, the interactions between bound electrons arising from correlation effects and exchange forces are significant whereas in the outer region, these effects become minimal and the electron trajectory is effectively determined by the long range potential exerted by the nuclei of the system.  The boundary between the inner and outer regimes is set to the maximum extent of the most diffuse orbital contained within the structure.  The $R$-matrix is defined as

\begin{equation} \label{R_Matrix_Boundary}
    R_{ij} = \frac{1}{2a} \sum^{N+1}_{k}\frac{\omega_{ik}(a)\omega_{jk}(a)}{E_{k}^{N+1}-E}
\end{equation}

where $a$ is the $R$-matrix boundary, $E_k^{N+1}$ are the eigenenergies of the ($N + 1$) Hamiltonian, $E$ is the energy of the incident electron, and the $\omega_{ik}$ are the energy independent surface amplitudes. The electron-impact excitation collision strength ($\Omega_{i\rightarrow j}$) between an initial state $i$ and final state $j$ is computed for all possible transitions. The relationship between the dimensionless collision strength and the cross section ($\sigma_{i\rightarrow j}$) is defined as

\begin{equation} \label{Collision_Strength}
    \Omega_{i \rightarrow j} = \frac{g_{i}k_{i}^{2}}{\pi a_{0}^2}\sigma_{i \rightarrow j}
\end{equation}

where $g_{i}$ is the statistical weight of the initial level, $k_{i}^2$ is the energy of the incident electron in Ryd, and $a_0$ is the Bohr radius. These collision strengths typically display some structure due to resonances converging on to the target state thresholds at low energies.  At higher energies the collision strength tends towards a high energy limit depending on the type of transition being represented.

Important for astrophysical modelling are the corresponding Maxwellian averaged effective collision strengths ($\Upsilon_{i\rightarrow j}$) where the averaging is performed over a Boltzmann distribution of electron temperatures. This is a much better approximation of the free electron distribution in many astrophysical and laboratory plasmas, where changes in plasma temperature result in slower variations of electron behaviour.  The conversion from collision to effective collision strength is described as 

\begin{equation} \label{Effective_Collision_Strength}
    \Upsilon_{i \rightarrow j}(T_e) = \int^{\infty}_{0}\Omega_{i \rightarrow j} e^{-\epsilon_{j}/kT_{e}}d\left( \frac{\epsilon_{j}}{kT_e} \right)
\end{equation}

where $\epsilon_j$ is the energy of the scattered electron, $k$ is the Boltzmann constant and $T_e$ is the electron temperature, in Kelvin.

Two variants of the $R$-matrix computer packages were used in this work. The Sr {\sc i} and Sr {\sc iv} AS model structures were incorporated into the semi-relativistic Breit-Pauli ({\sc rmbp}) suite of codes (\cite{badnell_1986}, \cite{badnell_1997}), where intermediate coupling is employed in the target wavefunction and the ($N + 1$) Hamiltonian includes the one-body correction operators only.    The Sr {\sc iii} and Sr {\sc iv} {\sc grasp}$^{0}$ structures formed the input into the fully relativistic Dirac Atomic $R$-matrix Code ({\sc darc}) suite where the Dirac Hamiltonian is solved within a fully relativistic $jj$ coupled scattering calculation  (\cite{Norrington_1987}).  Both the {\sc rmbp} and {\sc darc} coding suite may be found at \cite{Ballance_2024}. In the following subsections, we discuss the electron-impact excitation calculations for each Sr species and the relevant parameters used in their associated calculations. Sample collision and effective collision strengths will be shown for each ion with the full data sets being made available at \cite{Ballance_2024}, on the \cite{OPEN_ADAS} website or directly from the corresponding author.

\subsection{Electron-Impact Excitation of Sr {\sc i}} \label{sec:SrI-Excitation}

The 202 energy levels from the AS Sr {\sc i} target were reduced to the lowest lying 57 for the close-coupling scattering calculations and all were calibrated to their experimental positions listed in NIST.  A continuum orbital basis of 43 was selected, and the boundary between the inner and outer regions in the $R$-matrix calculations was 108.92au. The partial waves included in the scattering calculations were those where 1 $\leq 2J \leq$ 81 and two separate energy mesh grids were used.  A fine mesh grid (1 $\times10^{-4}$ Ryd) was applied for the partial waves 1 $\leq 2J \leq$ 39, consisting of 10,000 points covering the low energy region up to 1 Ryd, and a coarser mesh (2.5 $\times10^{-4}$ Ryd) for the higher partial waves with 41 $\leq 2J \leq$ 81. 
Contributions from partial waves with total angular momenta $2J >$ 81, of particular importance for the slowly converging allowed transitions, are represented using a 'top-up' procedure described in \cite{Burgess_1974} and \cite{Burke_1986}. 
To approximate the collision strengths at higher temperatures, an infinite energy point is also included using the procedure of \cite{Burgess_1992} for electric dipoles and \cite{eissner_1991} for all other transitions. A similar procedure will be adopted for all the Sr charge states considered in this work.  

\begin{figure*}
\centering
\includegraphics[width=\textwidth, height=9.5cm]{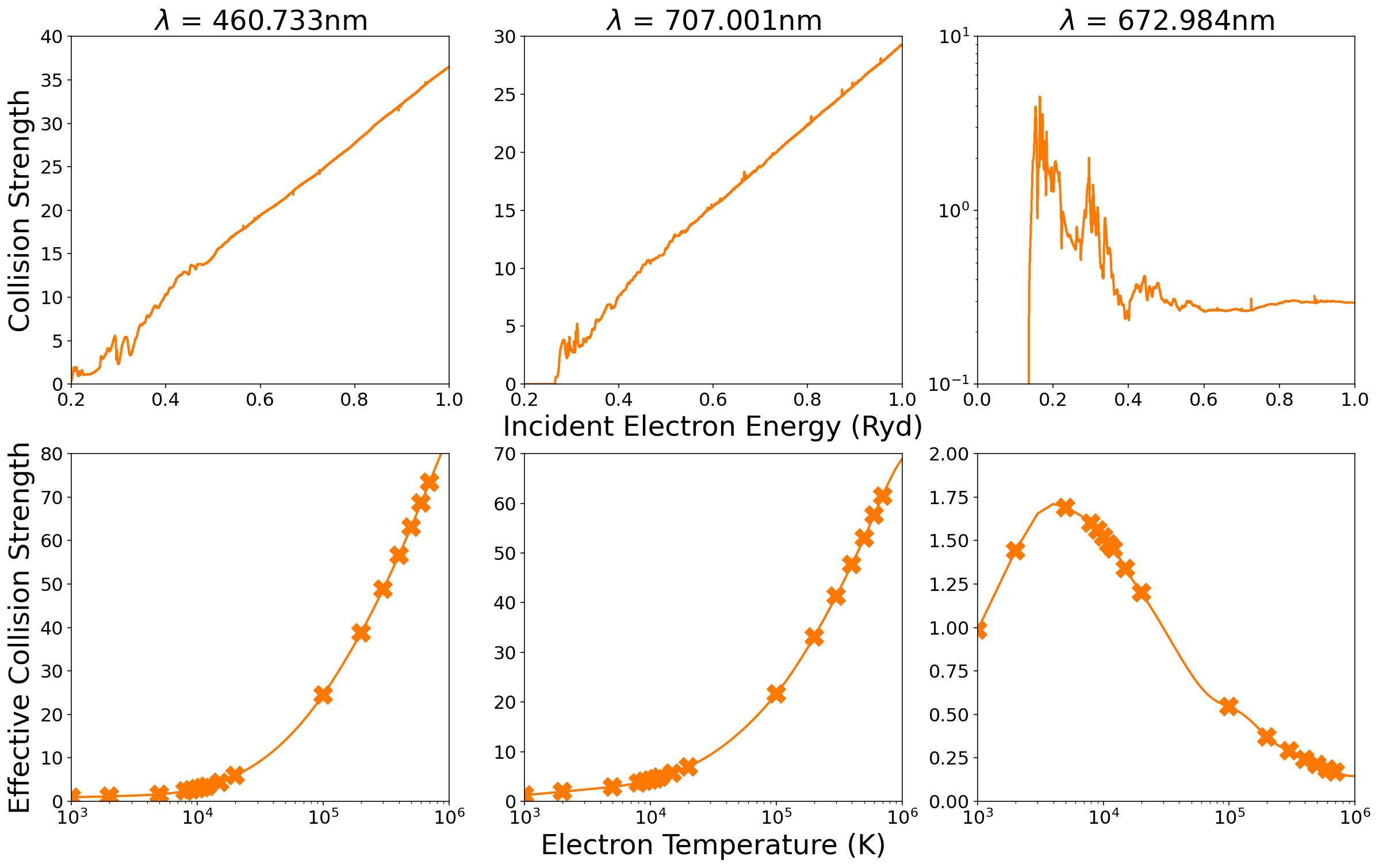}
\caption{Collision strengths (Top Row) and corresponding effective collision strengths (Bottom Row) for selected transitions in Sr {\sc i}. The transitions selected are 5s$^2$ $^1$S$_0$ $\rightarrow$ 5s5p $^1$P$_1^{\circ}$ (Left Column), 5s5p $^3$P$_2^{\circ}$ $\rightarrow$ 5s6s $^3$S$_1$ (Centre Column) and 5s$^2$ $^1$S$_0$  $\rightarrow$ 5s5p $^3$P$_2^{\circ}$ (Right Column), and have observed wavelengths of 460.733nm, 707.001nm and 672.984nm respectively.}
\label{fig:SrI_Col}
\end{figure*}

In Figure \ref{fig:SrI_Col}, we present the collision strength as a function of incident electron energy in Ryd, and the corresponding effective collision strength as a function of electron temperature in K, for three significant transitions in Sr {\sc i}, two E1 dipole allowed and one spin changing forbidden line. The 460.733nm and 707.001nm E1 dipole transitions presented here were signposted in \cite{bergemann_2012} as useful in determining the stellar abundances in FGK stars.  These correspond to the transitions 5s$^2$ $^1$S$_0$  $\rightarrow$ 5s5p $^1$P$_1^{\circ}$  and 5s5p $^3$P$_2^{\circ}$ $\rightarrow$ 5s6s $^3$S$_1$.  We also present data for the 672.984nm line, corresponding to the transition 5s$^2$ $^1$S$_0$ $\rightarrow$ 5s5p $^3$P$_2^{\circ}$, which is deemed to be a promising candidate transition for a Sr lattice atomic clock (e.g. \cite{Lu_2023}, \cite{Klusener2024}). At present there are no other available data in the literature with which to compare these results.

%
%\begin{figure*}[!hp]
%\centering
%\includegraphics[scale=0.30]{Figures/SrI_Col_9-1.png}
%\caption{The (a)- Collision and (b)- Effective Collision Strength for the (9-1) E1 transition in the Sr {\sc i} target.  This corresponds to the transition 5s5p $^1$P$_1^{\circ}$ $\rightarrow$ 5s$^2$ $^1$S$_0$.  The emission is observed at a wavelength of 460.7330nm.}
%\label{fig:SrI_Col1}
%\end{figure*}
%
%\begin{figure*}[!hp]
%\centering
%\includegraphics[scale=0.30]{Figures/SrI_Col_10-4.png}
%\caption{The (a)- Collision and (b)- Effective Collision Strength for the (10-4) E1 transition in the Sr {\sc i} target.  This corresponds to the transition 5s6s $^3$S$_1$ $\rightarrow$ 5s5p $^3$P$_2^{\circ}$.  The emission is observed at a wavelength of 707.0010nm.}
%\label{fig:SrI_Col2}
%\end{figure*}
%
%\begin{figure*}[!hp]
%\centering
%\includegraphics[scale=0.30]{Figures/SrI_Col_2-1.png}
%\caption{The (a)- Collision and (b)- Effective Collision Strength for the (2-1) M2 transition in the Sr {\sc i} target.  This corresponds to the transition 5s5p $^3$P$_0^{\circ}$ $\rightarrow$ 5s$^2$ $^1$S$_0$.  The emission is observed at a wavelength of 700.2950nm.}
%\label{fig:SrI_Col3}
%\end{figure*}
%
%\begin{figure*}[!hp]
%\centering
%\includegraphics[scale=0.30]{Figures/SrI_Col_4-1.png}
%\caption{The (a)- Collision and (b)- Effective Collision Strength for the (4-1) M2 transition in the Sr {\sc i} target.  This corresponds to the transition 5s5p $^3$P$_2^{\circ}$ $\rightarrow$ 5s$^2$ $^1$S$_0$.  The emission is observed at a wavelength of 672.9893nm.}
%\label{fig:SrI_Col4}
%\end{figure*}

\subsection{Electron-Impact Excitation of Sr {\sc iii}} \label{sec:SrIII-Excitation}

The 853 energy levels present in the initial Sr {\sc iii} target were reduced to the lowest 65 for the scattering calculations and all were shifted to their experimental positions listed in NIST.  A total of 40 continuum orbitals were included and the $R$-matrix boundary between the inner and outer regions set to 39.93au.  The close-coupling scattering calculations were performed for partial waves with total angular momentum $2J \leq 79$, using a fine energy mesh with a spacing of 5$\times10^{-5}$ Ryd adopted for the low $2J$ values and a coarser mesh of 1.5$\times10^{-4}$ Ryd for the higher $J$ values. 
These parameters allowed us to extend to energies up to $\sim$6.40 Ryd, which is considered a suitable temperature range for KNe and WD research.  

\begin{figure*}
\centering
\includegraphics[width=\textwidth, height=9.5cm]{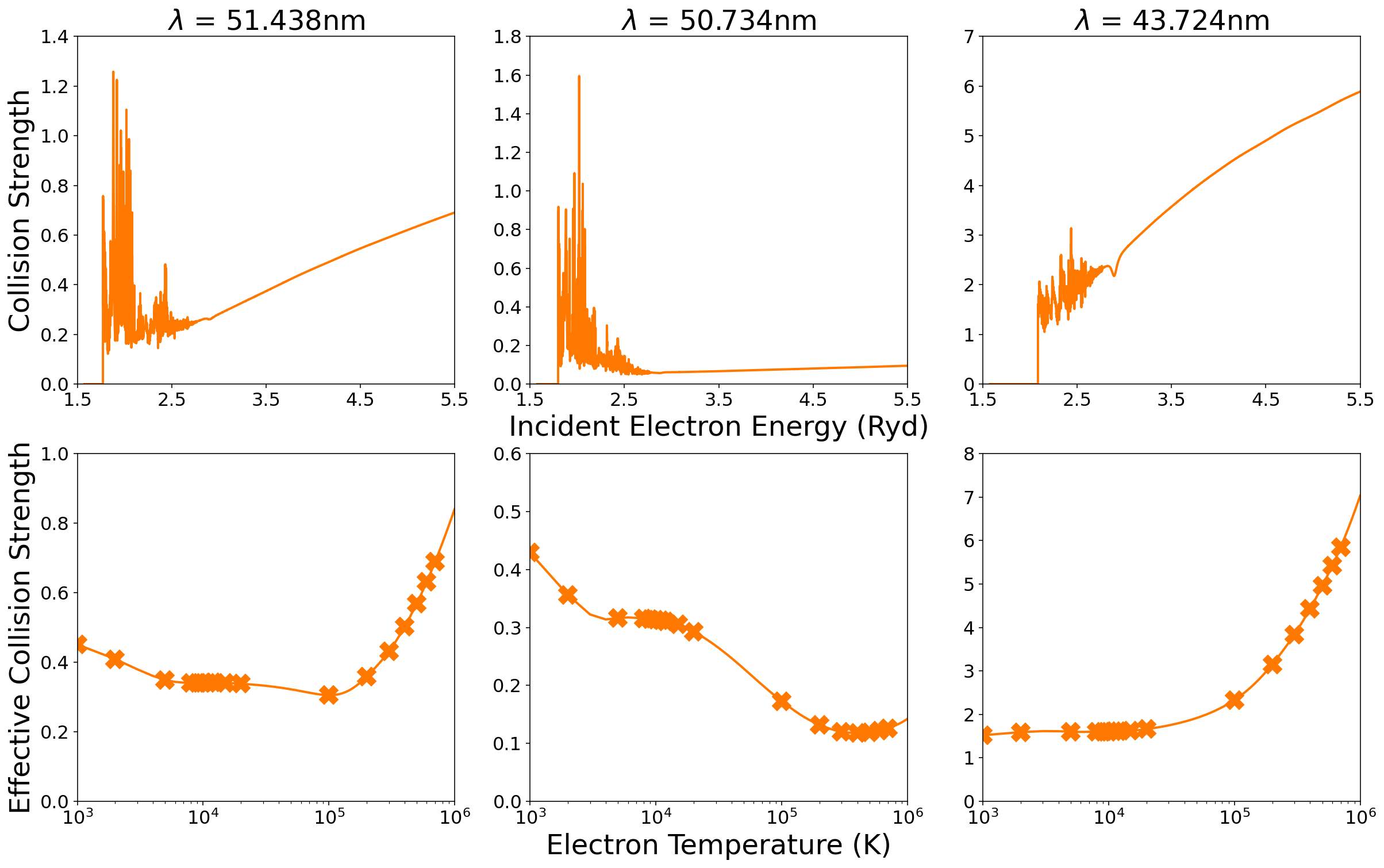}
\caption{The Collision (Top Row) and Effective Collision (Bottom Row) Strengths of selected transitions in our Sr {\sc iii} target.  The transitions selected are 4p$^6$ $^1$S$_0$ $\rightarrow$  4p$^5$5s $^1$P$_1^{\circ}$ (Left Column), 4p$^6$ $^1$S$_0$ $\rightarrow$  4p$^5$4d $^3$D$_1^{\circ}$ (Centre Column) and 4p$^6$ $^1$S$_0$ $\rightarrow$ 4p$^5$4d $^1$P$_1^{\circ}$ (Right Column), and have observed emissions at 51.438nm, 50.734nm and 43.724nm respectively.}
\label{fig:SrIII_Col}
\end{figure*}

For this species the selected sample of collision and effective collision strengths were for two E1 dipole lines and one intercombination line, as listed in \cite{sureau_1984}.  These are shown in Figure \ref{fig:SrIII_Col} and correspond to the E1 transitions 4p$^6$ $^1$S$_0$ $\rightarrow$  4p$^5$5s $^1$P$_1^{\circ}$ and 4p$^6$ $^1$S$_0$ $\rightarrow$  4p$^5$4d $^1$P$_1^{\circ}$, as well as the intercombination transition  4p$^6$ $^1$S$_0$ $\rightarrow$  4p$^5$4d $^3$D$_1^{\circ}$. There were no collision or effective collision strengths in the literature with which to compare our results.

\subsection{Electron-Impact Excitation of Sr {\sc iv}} \label{sec:SrIV-Excitation}

The 504 energy levels present in the Sr {\sc iv} were reduced to the lowest lying 70 energy levels, which were subsequently shifted to their NIST values for the close-coupling scattering calculations. A continuum orbital basis of 20 was assigned to each partial wave, with the $R$-matrix boundary between the inner and outer region set to 24.22au. The scattering calculations were performed for each partial wave where $2J \leq$ 68, with top up included for the partial waves where $2J \geq$ 68. The energy grid consisted of 24,000 evenly distributed points with a mesh size of 2.8 $\time10^{-5}$ Ryd. This spans the energy range of interest up to $\sim$6 Ryd.

\begin{figure*}
\centering
\includegraphics[width=\textwidth, height=9.5cm]{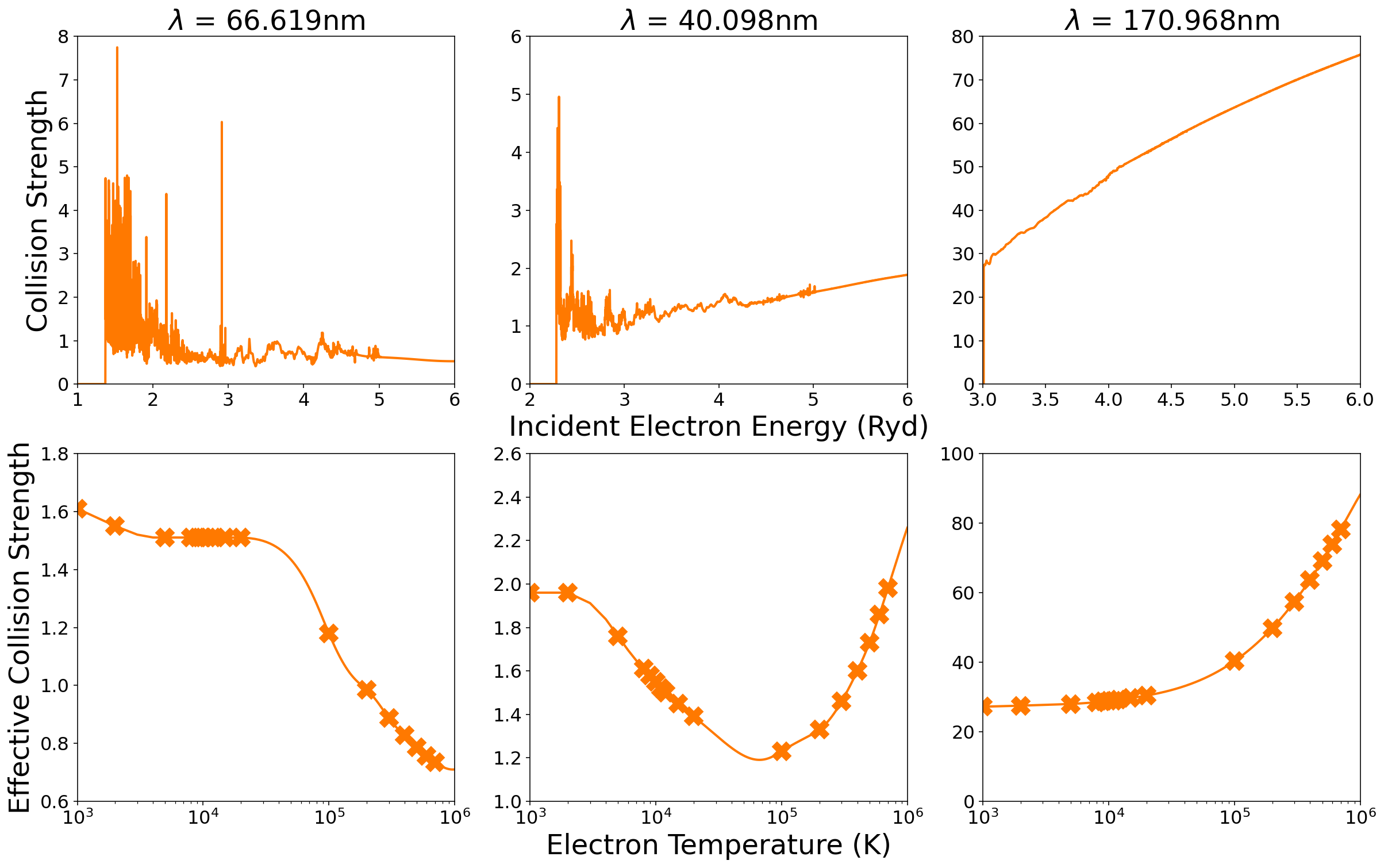}
\caption{The Collision (Top Row) and Effective Collision (Bottom Row) Strengths of selected transitions in our Sr {\sc iv} target.  The transitions selected are 4s$^2$4p$^5$ $^2$P$_{3/2}^{\circ}$ $\rightarrow$ 4s4p$^6$ $^2$S$_{1/2}$ (Left Column), 4s$^2$4p$^5$ $^2$P$_{3/2}^{\circ}$ $\rightarrow$  4s$^2$4p$^4$5s $^2$D$_{5/2}$ (Centre Column) and 4s$^2$4p$^4$5p $^2$D$_{5/2}^{\circ}$ $\rightarrow$  4s$^2$4p$^4$5d $^2$F$_{7/2}$ (Right Column), and have observed emissions at 66.916nm, 40.098nm and 170.968nm respectively.}
\label{fig:SrIV_Col}
\end{figure*}

In Figure \ref{fig:SrIV_Col} we present a representative sample of collision and effective collision strengths for some of the strong E1 transitions in the Sr {\sc iv} target.  These lines correspond to emission wavelengths of 66.6103nm, 40.0984nm and 170.9679nm and relate to transitions 4s$^2$4p$^5$ $^2$P$_{3/2}^{\circ}$ $\rightarrow$ 4s4p$^6$ $^2$S$_{1/2}$,  4s$^2$4p$^5$ $^2$P$_{3/2}^{\circ}$ $\rightarrow$ 4s$^2$4p$^4$5s $^2$D$_{5/2}$ and 4s$^2$4p$^4$5p $^2$D$_{5/2}^{\circ}$ $\rightarrow$ 4s$^2$4p$^4$5d $^2$F$_{7/2}$ respectively.   There were no corresponding collision strengths in the literature for comparison.

\subsection{Electron-Impact Excitation of Sr {\sc v}} \label{sec:SrV-Excitation}

Finally, the 1817 energy levels for the Sr {\sc v} calculations were reduced to the first shifted 95 energy levels.  An orbital continuum basis of 25 was allocated to each partial wave, which were generated for total angular momenta corresponding to 1 $\leq 2J \leq$ 91. The R-matrix boundary between the inner and outer regions was set to 26.01au and the collision strengths were computed for a fine mesh (1.7 $\times10^{-5}$ Ryd) of energies to properly delineate the resonance structures. 

\begin{figure*}
\centering
\includegraphics[width=\textwidth, height=9.5cm]{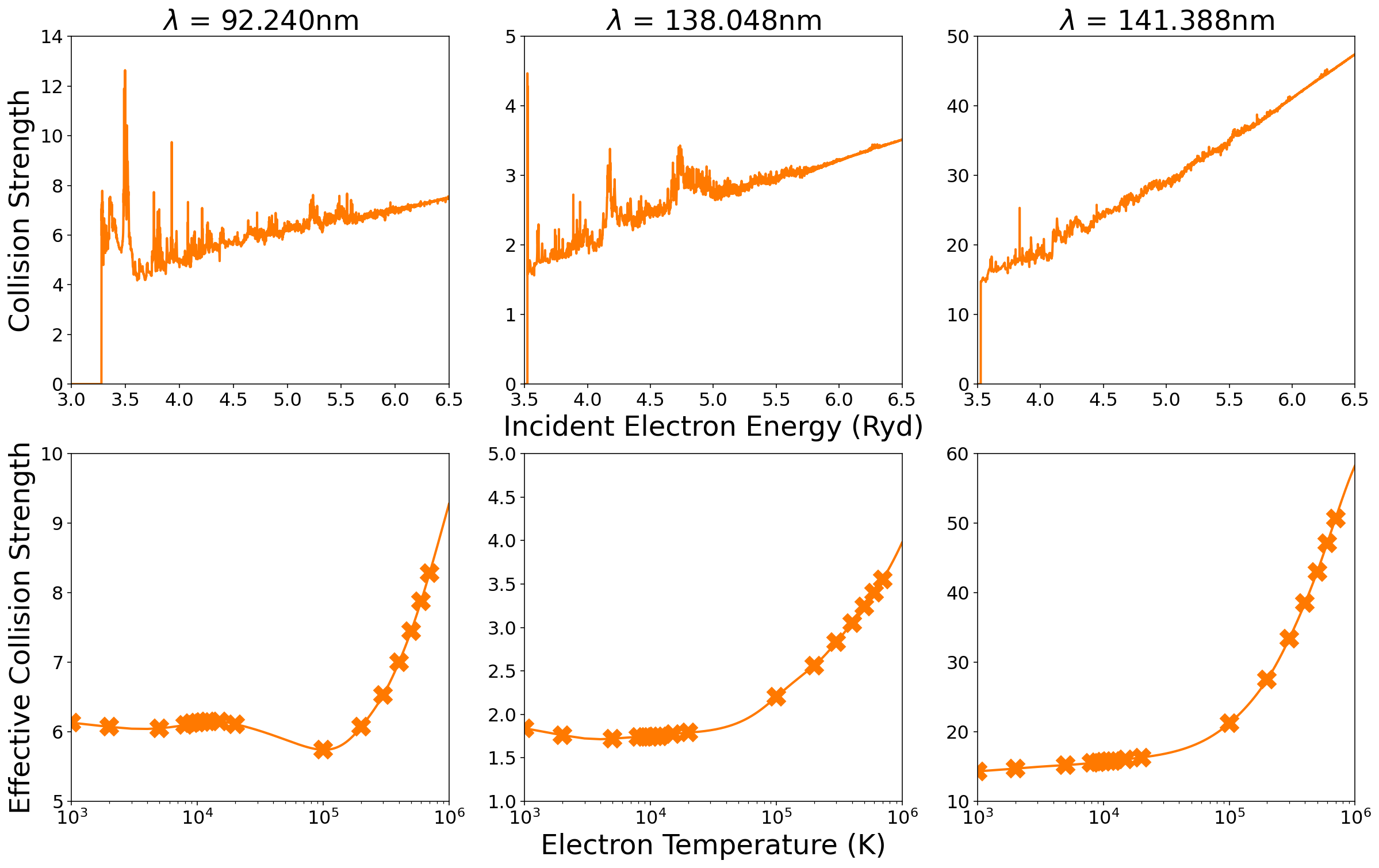}
\caption{The Collision (Top Row) and Effective Collision (Bottom Row) Strengths of selected transitions in our Sr {\sc v} target.  The transitions selected are 4s$^2$4p$^3$4d $^3$F$_4^{\circ}$ $\rightarrow$ 4s$^2$4p$^3$5p $^3$D$_3$ (Left Column), 4s$^2$4p$^3$5p $^5$P$_3$ $\rightarrow$  4s$^2$4p$^3$5d $^5$D$_2^{\circ}$ (Centre Column) and 4s$^2$4p$^3$5p $^5$P$_3$ $\rightarrow$ 4s$^2$4p$^3$5d $^5$D$_4^{\circ}$ (Right Column), and have observed emissions at 101.371nm, 138.048nm and 141.388nm respectively.}
\label{fig:SrV_Col}
\end{figure*}

In Figure \ref{fig:SrV_Col} we present the collision and effective collision strengths for three sample transitions in Sr {\sc v}. These lines were chosen as they were identified in \cite{Rauch_2017} as important in the modelling of the hot WD RE 0503-289 at wavelengths 92.2397nm, 138.0477nm and 141.3882nm respectively. These correspond to the transitions 4s$^2$4p$^3$4d $^3$F$_4^{\circ}$ $\rightarrow$ 4s$^2$4p$^3$5p $^3$D$_3$, 4s$^2$4p$^3$5p $^5$P$_3$ $\rightarrow$ 4s$^2$4p$^3$5d $^5$D$_2^{\circ}$ and 4s$^2$4p$^3$5p $^5$P$_3$ $\rightarrow$ 4s$^2$4p$^3$5d $^5$D$_4^{\circ}$. Again there are no corresponding collision strengths in the literature for comparison.

\section{Collisional Radiative Modelling under NLTE Conditions}
\label{sec:Colisional Radiative Modelling}

To simulate the NLTE collisional radiative processes within an astrophysical plasma, we employed the {\sc colradpy} Python package described in \cite{Johnson_2019}.  This is based on the collisional radiative theory outlined in \cite{Bates_1962} and later generalised in \cite{Summers_2006}.  The energy levels, quantum numbers, A-values and effective collision strengths presented in the earlier sections serve as the input.  For each level $N_i$, {\sc colradpy} solves the following differential equation.

\begin{equation}
    \frac{dN_j}{dt} = \sum_iC_{ij}N_i
\end{equation}

where $C_{ij}$ is the collisional matrix representing the excitation/de-excitation processes corresponding to levels $i$ and $j$ in the target structure. This package can be run for a wide range of electron temperature and density parameters.  In this work, we prioritised two specific ranges. Firstly, we investigate the KNe regime for temperatures between 0.05eV $\leq T_e \leq$ 1.00eV and an electron density covering 1x10$^6$cm$^{-3} \leq n_e \leq$ 1x10$^9$cm$^{-3}$.  This is similar to the parameter space employed in other KNe related investigations, such as the modelling work in \cite{Gillanders_2023} and \cite{mulholland_2024}.  Secondly, we probe conditions relevant to those observed in a WD with particular attention focused on the WD star RE 0503-289, where the Sr {\sc v} lines were identified in its spectra by \cite{Rauch_2017}, and was discerned in \cite{Barstow_1994} to have an effective surface temperature of 70000K ($\sim$6eV).  For the WD regime, we adopt a temperature grid of 4eV $\leq T_e \leq$ 8eV and a density grid of 1x10$^{16}$cm$^{-3} \leq n_e \leq$ 1x10$^{20}$cm$^{-3}$.  In addition, to simulate LTE conditions we set the plasma density to an extremely high value to facilitate comparison, and also to investigate the transition from LTE to NLTE conditions.  In this analysis, and for completeness, the atomic data for Sr {\sc ii} computed in \cite{mulholland_2024} will also be used to allow for a full analysis of the collisional radiative processes for all low-lying ion stages of Sr. 

\subsection{Synthetic Spectra Emissions}
\label{sec:PEC}

\begin{figure*}
\includegraphics[scale=0.40]{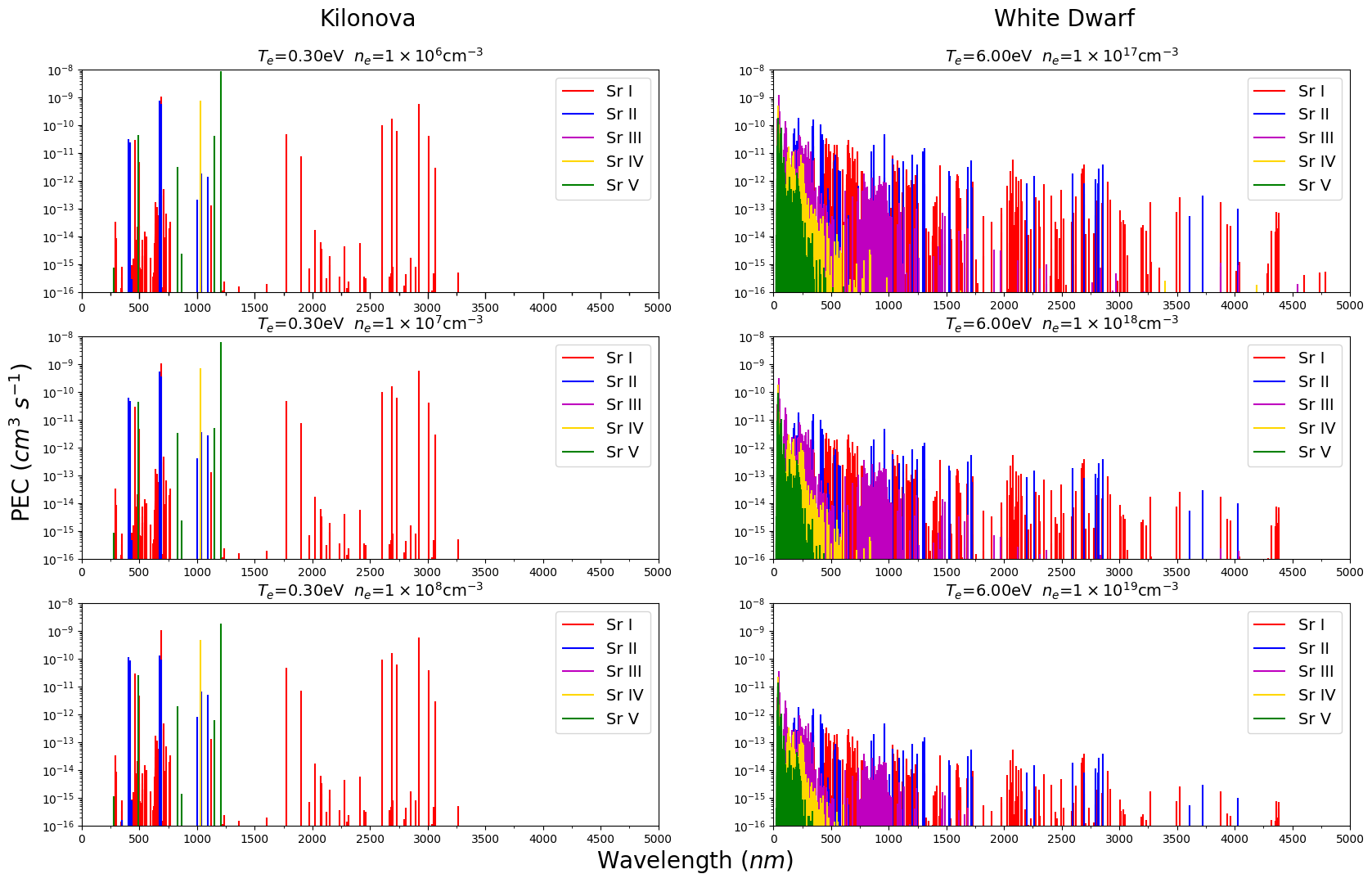}
\caption{The Photon Emissivity Coefficients (PECs) for the first five ionisation stages of Sr for a plasma of electron temperatures and densities relevant to KNe (left panel) and WD (right panel) events in the wavelength window of 0 - 5000nm.}
\label{fig:PEC_Both}
\end{figure*}

To simulate the emission spectrum of a Sr source, an analysis of the Photon Emissivity Coefficients (PECs) is carried out.  Each PEC corresponds to one transition within an ion species, and can be determined as

\begin{equation}
    PEC^{exc}_{j\rightarrow i} = A_{j\rightarrow i} \frac{N_j}{n_e}    
\end{equation}

where $N_j$ is the population of the upper level weighted by the groundstate population and $A_{j\rightarrow i}$ is the corresponding Einstein A coefficient. It is noted that the populations are carried out independently for each ion stage and therefore the relative heights of spectral features between ion stages have no physical meaning.  Only the relative heights within an ion stage have significance.  PECs are typically expressed in units of cm$^3$s$^{-1}$, where a PEC of a higher magnitude indicates that the emission line corresponding to the transition is likely to be more prominent in the observed spectra of the source.  In \text{Figure \ref{fig:PEC_Both}} we present sample PEC plots for the first five ionisation stages over the wavelength window of 0 - 5000nm, under both KNe (left panel) and WD (right panel) regimes.  The temperature of the plasma was set to 0.3eV and 6.0eV for the KNe and WD plasmas respectively.

\begin{figure}
\includegraphics[scale=0.30]{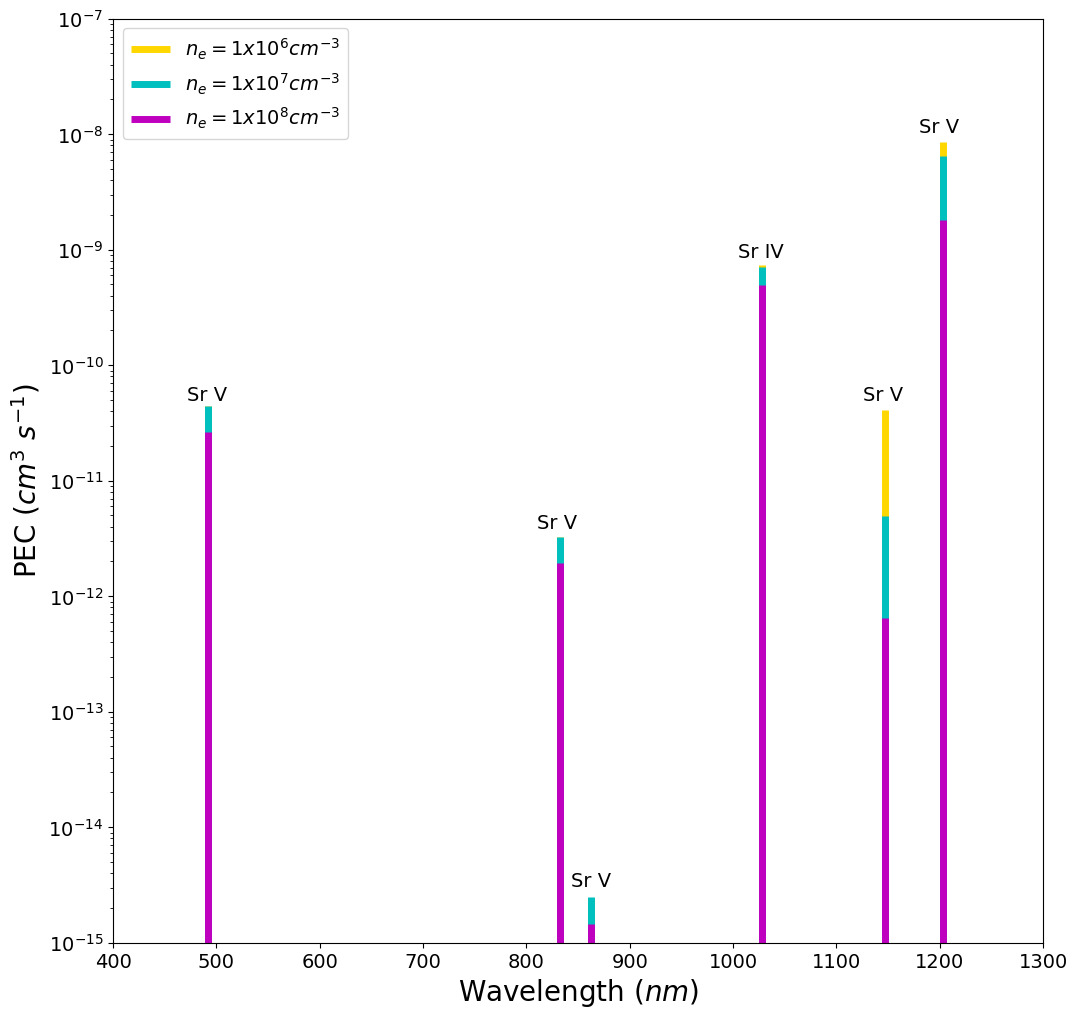}
\centering
\caption{The Photon Emissivity Coefficient (PECs) for Sr {\sc iv} and {\sc v} over a wavelength range of 400nm to 1300nm at a temperature of 0.30eV at different electron density parameters.  The PEC corresponding to each ion is highlighted above each line.}
\label{fig:PEC_Kilonoave_High_Ion}
\end{figure}

In the KNe regime, the spectrum is dominated by Sr {\sc i} emissions, with a number of Sr {\sc ii} lines present at shorter wavelengths. Particularly clear are the three Sr {\sc ii} lines between 1000 and 1100nm, corresponding to the wavelength region where the well identified P Cygni spectral line of Sr {\sc ii}
at approximately 1$\mu$m is located (\cite{watson_2019}, \cite{Gillanders_2023}).  
Also evident are strong Sr {\sc ii} emission lines at $\sim 400$ and $650$ nm. These are due to the low-lying transitions among the first five levels (1-5, 1-4, 1-3, 1-2) at 407.89, 421.67, 674.03 and 687.01nm respectively, as discussed in \cite{mulholland_2024}. 

There are no Sr {\sc iii} emission lines present. This is because the first excited level of Sr {\sc iii} has an energy of $\sim$1.60 Ryd (\cite{Persson_1972}, \cite{Hansen_1973}), significantly higher than the temperatures observed in the spectra of KNe events. In addition, the ground configuration of Sr {\sc iii} is comparable to that of a noble gas, and as such there is only one level associated with it.  This means that, if present, the abundance of Sr {\sc iii} will be mostly restricted to its ground state and thus no emission would be expected to be observed.

Emission features from the higher ionisation stages of Sr {\sc iv} and {\sc v} are present in \text{Figure \ref{fig:PEC_Both}} at short wavelengths below 1300nm but are limited to a small number of six lines. To isolate these lines for further investigation we plot in \text{Figure \ref{fig:PEC_Kilonoave_High_Ion}} the PECs for these species over the small wavelength range from 400 -  1300nm and for a selection of densities (10$^{6}$, 10$^{7}$, 10$^{8}$cm$^{-3}$). There is a single Sr {\sc iv} line at 1027.69nm (1-2), and five Sr {\sc v} lines at 492.22nm (1-4), 832.92nm (2-4), 862.42nm (3-4), 1146.69nm (1-3) and 1203.35nm (1-2) presented.  These correspond to forbidden transitions between the levels which make up the ground configurations of Sr {\sc iv} (4s$^2$4p$^5$) and Sr {\sc v} (4s$^2$4p$^4$). The relative PEC strengths are comparable, and in the particular case of the 1203.35nm Sr {\sc v} line, stronger than the other Sr {\sc i} and Sr {\sc ii} lines presented in the full spectra of \text{Figure \ref{fig:PEC_Both}}.  In addition, the PEC strengths are shown to be responsive to varying electron densities within the KNe regime, especially for the Sr {\sc iv} line at 1027.69nm and the strong Sr {\sc v} line at 1203.35nm.

Higher charge states of Sr have yet to be identified in KNe spectra.  The energies required to ionise Sr {\sc i} to three and four times ionised ($\sim$3-4 Ryd) means that there will likely be only a trace amount of Sr {\sc iv} and {\sc v} present in the event as neutral and singly ionised Sr will dominate.   However, the work of \cite{CarvajalGallego_2024} shows that the presence of highly ionised lanthanide ions can have an effect on the observed KNe spectra, and as such it is possible for the Sr {\sc iv} and {\sc v} ions to have a similar effect. The Sr {\sc v} 1-2 transition at 1203.35nm could be a suitable candidate to test this due to its strong PEC coefficient that is sensitive to density variations.

In the WD conditions (right panel of \text{Figure \ref{fig:PEC_Both}}), the spectrum is significantly more populated.  The emission lines corresponding to all five ionisation stages of Sr are present.  At shorter wavelengths (<1,000nm), the region is overcrowded with lines of similar magnitude from all five ion stages. At longer wavelengths, the spectra becomes more sparsely populated and is dominated by Sr {\sc i} and {\sc ii} features, with no strong Sr {\sc iii} - {\sc v} lines evident. The PECs are susceptible to degrees of variation of the electron densities expected for WDs.  As the shorter wavelengths are highly populated, the infrared region could be a better site to probe for potential temperature and density sensitive diagnostic lines.  In particular, the 3000 -  4000nm wavelength window is particularly interesting as  the Sr {\sc i} and {\sc ii} lines are unblended and the corresponding PECs are of a suitably large magnitude to investigate diagnostic lines in WD stars.   

\subsection{Level Population Analysis for the Identification of Temperature and Density Diagnostics}
\label{sec:Populations}

\begin{figure}
\centering
\includegraphics[scale=0.44]{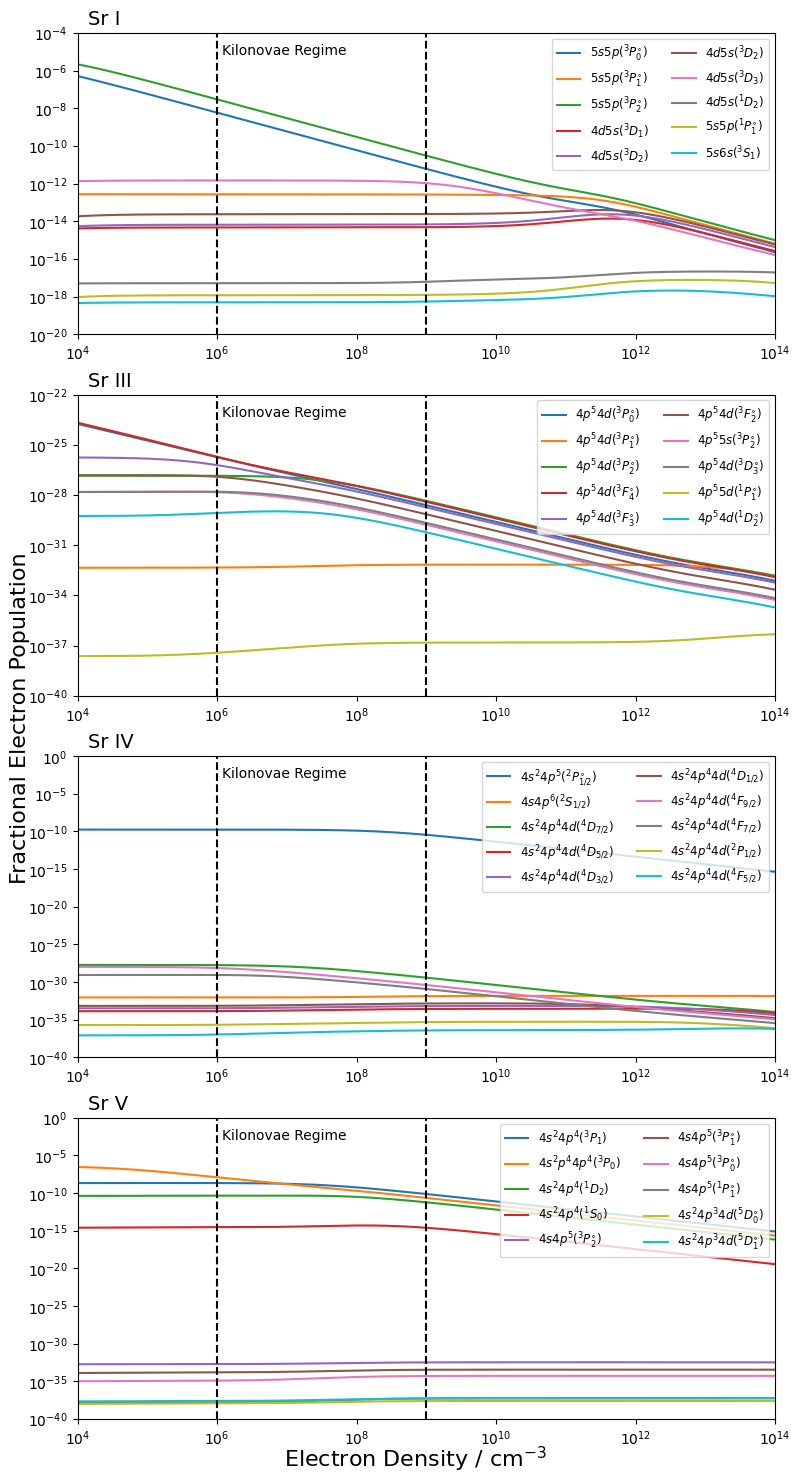}
\caption{The variations in the fractional population of the first 10 levels of Sr {\sc i}, {\sc iii}, {\sc iv} and {\sc v} with electron temperature of the plasma at $T_e=$0.50eV.  The region applicable to KNe events is highlighted between the two black, dashed, vertical lines.}
\label{fig:Pop_Kilonoave}
\end{figure}

\begin{figure}
\centering
\includegraphics[scale=0.44]{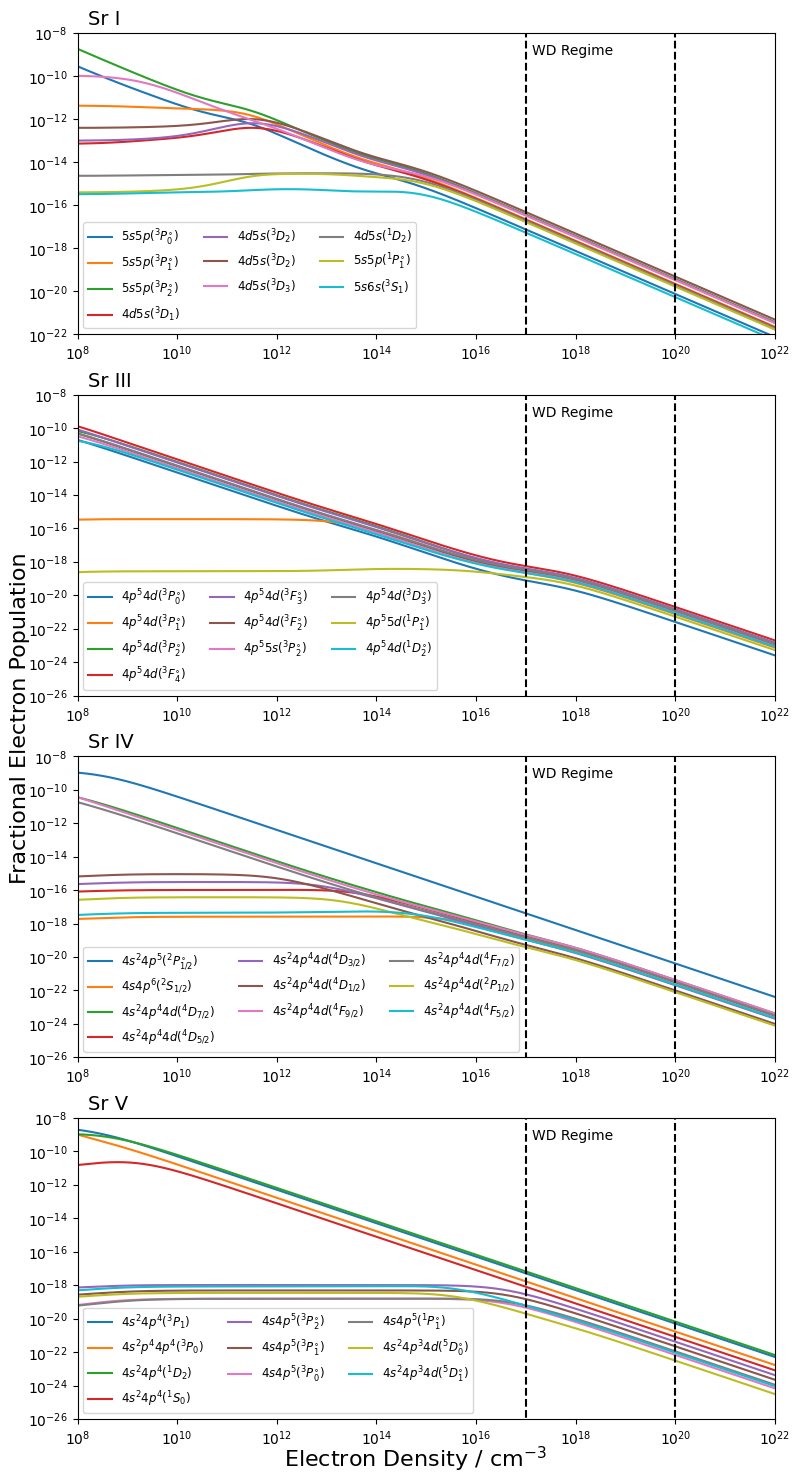}
\caption{The variations in the fractional populations for the first 10 levels of Sr {\sc i}, {\sc iii}, {\sc iv} and {\sc v} with electron temperature of the plasma at $T_e$=6.00eV.  The region applicable for WD conditions is highlighted between the two black, dashed, vertical lines.}
\label{fig:Pop_WD}
\end{figure}

\begin{figure*}
\centering
\includegraphics[scale=0.50]{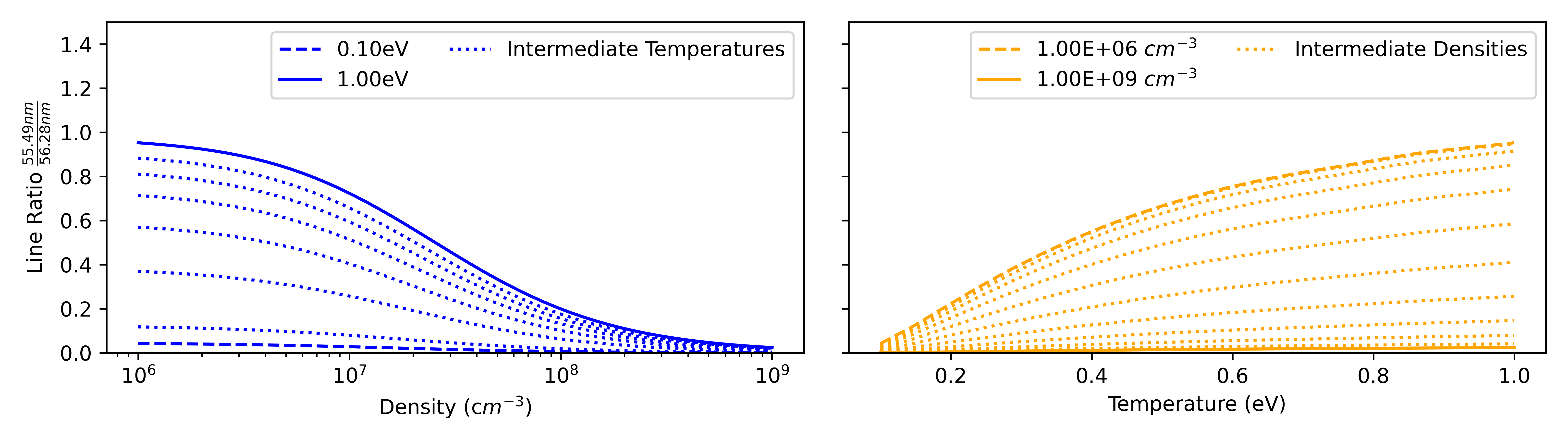}
\caption{Sample density (left) and temperature (right) diagnostic line ratios for Sr {\sc iii} under KNe conditions.  The line ratio employed was $\frac{1-4}{1-3}$, which corresponds to the transitions 4p$^6$ $^1$S$_{0}$ $\rightarrow$ 4p$^5$4d $^3$P$_2^{\circ}$ ($\lambda = 55.49$nm) and  4p$^6$ $^1$S$_{0}$ $\rightarrow$ 4p$^5$4d $^3$P$_1^{\circ}$ ($\lambda = 56.28$nm). }
\label{fig:Diagnostic_Kilonoave}
\end{figure*}

\begin{figure*}
\centering
\includegraphics[scale=0.50]{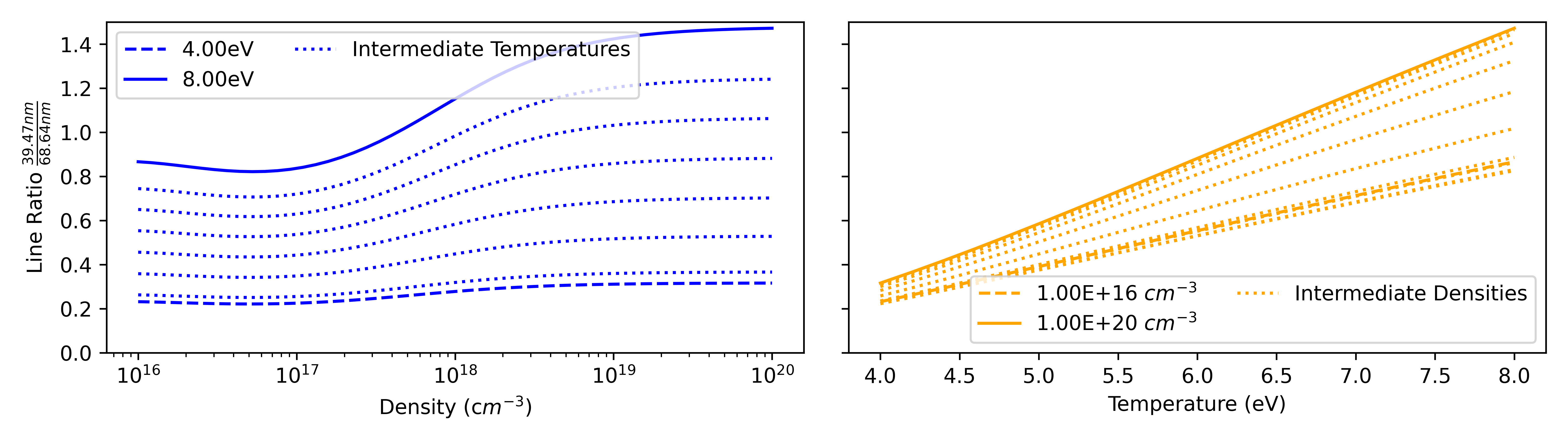}
\caption{Sample density (left) and temperature (right) diagnostic lines for Sr {\sc v} under WD conditions.  The line ratio employed was $\frac{1-34}{2-6}$, which corresponds to the transitions 4p$^4$ $^3$P$_2$ $\rightarrow$ 4p$^3$4d $^3$D$_3^{\circ}$ ($\lambda=39.47$nm) and 4s$^2$4p$^4$ $^3$P$_1$ $\rightarrow$ 4s4p$^5$ $^3$P$_2^{\circ}$ ($\lambda = 68.64$nm).}
\label{fig:Diagnostic_WD}
\end{figure*}

The quality of a diagnostic line, whether it be density or temperature, depends upon the ratio of the two constituent transitions.  The ratio needs to exhibit variations in electron density and/or temperature to qualify as a plasma diagnostic.  To identify possible diagnostic lines for Sr {\sc i}, {\sc iii}, {\sc iv} and {\sc v}, we initially performed an analysis of the electron populations of each level retained in the Sr models using the collisional radiative modelling code {\sc colradpy} (\cite{Johnson_2019}). 

In \text{Figure \ref{fig:Pop_Kilonoave}} and \text{Figure \ref{fig:Pop_WD}}, we present the fractional electron populations as a function of electron density in units cm$^{-3}$ for the first 10 levels of Sr {\sc i}, {\sc iii}, {\sc iv} and {\sc v} at temperatures relevant to KNe (T$_{e}$ = 0.5eV, \text{Figure \ref{fig:Pop_Kilonoave}}) and WD (T$_{e}$ = 6.0eV, \text{Figure \ref{fig:Pop_WD}}) related plasmas. In each figure we have highlighted with dashed black lines the density range most applicable to KNe and WD modelling applications. It should be noted that while only the first 10 levels are shown, we checked the behaviours of the remaining excited levels and the populations for the higher excited states followed similar trends.  The level populations for each ion stage exhibit three distinct regions of behaviour.  The first transpires when the population remains constant with varying electron density. This occurs at lower electron densities and suggests that the rates of collisional excitation and spontaneous emission are of similar magnitudes, and hence there is no net gain or loss of electrons from this level.  In other words, a coronal approximation is valid here.  The second trend occurs at higher densities, where there is a negative linear dependence on the electron population with density. This indicates the collisional processes are dominant in the system, exciting the electrons out of the levels at a rate faster than which electrons are entering the system through radiative spontaneous emission.  Essentially, LTE conditions apply and the level populations can be represented using a Saha-Boltzmann distribution curve, given as

\begin{equation}
   \frac{N_i}{N_+} = n_e\left( \frac{h^2}{2\pi mkT_e}\right)^{\frac{3}{2}} \frac{g_i}{2g_+}e^{\frac{I_i}{kT_e}}  
\end{equation}

where $N_i$ is the weighted population of the excited level, $N_+$ is the weighted population of the ion in the plasma, $h$ is the Planck constant, $g_i$ is the statistical weight of the excited level, $g_+$ is the statistical weight of the ion in the plasma, $m$ is the particle mass, and $I_i$ is the ionisation potential from the excited level.
The third region depicts the NLTE regime, the transition phase between the coronal and LTE regimes. There is a non-linear dependence of the electron populations as a function of density.
The electron population in each level will follow the transition from coronal to NLTE to LTE as the electron density and temperature increase.  The conditions upon which these phase transitions will occur are different for each level, and this can result in the electron populations displaying differing behaviours at particular temperatures and densities.

In the KNe regime (\text{Figure \ref{fig:Pop_Kilonoave}}), all three population behaviours appear to be present within the relevant electron density range 10$^{6}$ - 10$^{9}$ cm $^{-3}$.  For Sr {\sc i}, levels 1 and 3 exhibit LTE behaviour while the remaining eight levels follow the coronal approximation.  For Sr {\sc iii}, there is a mix of conditions present, especially at around densities of 10$^7$ cm$^{-3}$, where levels 3, 7, 9 and 10 transition from coronal to LTE behaviour.  For Sr {\sc iv}, most of the levels follow the coronal approximation, though level 1 does exhibit a degree of NLTE behaviour at the higher electron density limit of 10$^9$ cm$^{-3}$.  In addition, levels 3, 7 and 8 exhibit NLTE behaviour at around 10$^7$ cm$^{-3}$.  The levels in Sr {\sc v} are similar to those in Sr {\sc iv} in that the majority of them follow the coronal approximation, though level 2 is already following LTE conditions within the KNe regime.  Levels 1, 3 and 4 follow the transition into LTE conditions at the higher electron density limit. In contrast, for the high densities relevant to WD modelling (10$^{17}$ - 10$^{20}$ cm$^{-3}$ ), depicted in \text{Figure \ref{fig:Pop_WD}}, all levels typically exhibit the negative linear behaviour of LTE as expected.  There is very little variation in the trends of the level populations at such high density and temperature parameters.  

This population analysis for ground and excited levels of Sr species allow us to probe for potential temperature and density sensitive diagnostic lines. For KNe modelling we focus on the lower ion stages. In \text{Figure \ref{fig:Diagnostic_Kilonoave}}, we highlight one possible Sr {\sc iii} temperature and density sensitive diagnostic line, the $\frac{1-4}{1-3}$ line ratio with wavelengths of 55.49nm and 56.28nm respectively. A variety of plasma temperatures ranging from 0.10eV to 1.0eV are considered across the density range 10$^{6}$ $\leq$ n$_{e}$ $\leq$ 10$^{9}$ cm $^{-3}$, relevant for KNe modelling. 

For WD modelling, the majority of the level populations were either at or approaching LTE and hence the search for potential diagnostic lines was difficult. The density diagnostics were only valid over the lower limit of the WD electron density (n$_e$=10$^{17}$cm$^{-3}$).  In \text{Figure \ref{fig:Diagnostic_WD}}, we highlight one possible Sr {\sc v} diagnostic line.  The temperature range considered was 4.0eV $\leq T_e \leq$ 8.0eV, with an electron density range of 10$^{16}$cm$^{-3}$ $\leq$ n$_e$ $\leq$ 10$^{20}$cm$^{-3}$.  This line ratio is for the $\frac{1-34}{2-6}$ transitions in Sr {\sc v}, for wavelengths at 39.47nm and 68.64nm respectively.  This exhibits some density and temperature variation which could be useful for experimental work.

\section{Conclusions}
\label{sec:Conclusions}

Emission lines from the low ionisation stages of Sr are observed across a wide array of different astrophysical phenomena, from KNe events to the evolution of distant stars and galaxies.  
%(cut :  just a repeat of the intro They also provide promising candidates for the development of atomic lattice clocks.)  
We have aimed to provide accurate atomic data sets for energy levels, transition probabilities and excitation/de-excitation rates for these low charge stages of Sr, for use in the simulation and modelling of astrophysical phenomena.

Four atomic structure models, for Sr {\sc i}, {\sc iii}, {\sc iv} and {\sc v}, were developed using the {\sc grasp}$^0$ and {\sc autostructure} packages. The computed energy levels and Einstein A-values were found to be in good agreement with the experimental work presented in NIST, and other previously published work. Collisional calculations using the {\sc darc} and {\sc bprm} codes were subsequently performed to determine electron impact-excitation collision strengths and Maxwellian averaged effective collision strengths for all possible forbidden and allowed transitions included in the wavefunction representations of the Sr models considered. To our knowledge, no other data was available in the literature with which to compare these results.

Using the collisional radiative solver {\sc colradpy}, synthetic spectra comprising the first five ionisation stages of Sr were constructed, for both KNe and WD conditions. It was found that in the KNe regime the resulting spectra were comprised mostly of Sr {\sc i} and Sr {\sc ii} lines, although a single strong Sr {\sc iv} line and an additional five strong Sr {\sc v} lines were found to be present between 400 - 1300nm.  In particular, the Sr {\sc v} 1-2 transition at 1203.35nm was particularly strong and exhibited some variation as the electron density conditions were altered. The WD synthetic spectra comprised all five ionised species of Sr and the emission spectrum was shown to be densely populated at shorter wavelengths, with some Sr {\sc i} and {\sc ii} lines evident in the infrared wavelength window.

A further analysis of the electron level populations of the ground and excited states found that the majority of the levels for all Sr species followed either the LTE or coronal approximation for conditions similar to those in KNe and WD events. A subsequent probe for potential density and/or temperature sensitive diagnostic line ratios found only minimal candidates, the Sr {\sc iii} $\frac{1-4}{1-3}$ for KNe applications and the Sr {\sc v} $\frac{1-34}{2-6}$ line ratio for WDs. These diagnostic line ratios are useful when diagnosing experimental plasmas under a very narrow electron temperature and density parameter space. 

%Further work needs to be undertaken to complete the atomic data available for the strontium species.  In particular, a comprehensive data set for the other transition rates such as photoionisation, recombination and electron-impact ionisation have yet to be discerned, outside of a few select cases (e.g. \cite{bergemann_2012}, \cite{fernández-menchero_2020}).  These rates would also be needed to compliment the electron-excitation collision strengths derived in this work to further improve upon the initial data used in simulation codes such as TARDIS.  Through the modelling of distant astrophysical phenomena such as kilonovae and white dwarf stars, we can gain a better understanding of their specific reaction pathways.

\section*{Acknowledgements}

We are grateful for the use of computing resources from the Northern Ireland High Performance Computing (NI-HPC) service funded by EPSRC, United Kingdom (EP/T022175).  DJD thanks the Science and Technology Facilities Council (STFC) of the UK Research and Innovation (UKRI) body for their support through his studentship.

%The TOSS service (\hyperlink{http://dc.g-vo.org/TOSS}{http://dc.g-vo.org/TOSS}) used for this work was constructed as part of the activities of the German Astrophysical Virtual Observatory.

%%%%%%%%%%%%%%%%%%%%%%%%%%%%%%%%%%%%%%%%%%%%%%%%%%
\section*{Data Availability}

The effective collision strengths for Sr {\sc i}, Sr {\sc iii}, Sr {\sc iv} and Sr {\sc v} will be available on the following websites \cite{Ballance_2024} and open-adas at \url{https://open.adas.ac.uk/} in adf04 file format. The collision strengths can be shared upon request to the corresponding author.



%\begin{thebibliography}{99}
%\end{thebibliography}

%%%%%%%%%%%%%%%%%%%%%%%%%%%%%%%%%%%%%%%%%%%%%%%%%%

%%%%%%%%%%%%%%%%% APPENDICES %%%%%%%%%%%%%%%%%%%%%

%%%%%%%%%%%%%%%%%%%%%%%%%%%%%%%%%%%%%%%%%%%%%%%%%%

% Don't change these lines
\bsp	% typesetting comment
\label{lastpage}
\end{document}